\documentclass[useAMS,usenatbib]{mn2e}
\usepackage{epsfig,natbib}
\usepackage{amssymb}
\usepackage{graphicx}
\usepackage{placeins}
\usepackage{times}
\usepackage{epsfig}
\usepackage{amsmath}

\newcommand\ion[2]{#1{\sc #2}}%  ion, i.e. CII = \ion{C}{ii}
\newcommand{\Msun}{\ifmmode{M_\odot}\else$M_\odot$~\fi}

\newcommand{\kmsMpc}{\hbox{km\thinspace s$^{-1}$\thinspace Mpc$^{-1}$}} 
\renewcommand*{\vec}[1]{\mathbf{#1}}
\title[The DLA towards Q\,0918+1636 - a challenge to galaxy formation models?] 
{The galaxy counterpart of the high-metallicity and 16 kpc impact 
parameter DLA towards Q\,0918+1636 - a challenge to galaxy formation models?}
\author[Sommer-Larsen \& Fynbo]{
J. Sommer-Larsen$^{1,2,3}$\thanks{E-mail:jslarsen@astro.ku.dk} \&
J. P. U. Fynbo$^{1}$
\\
$^{1}$Dark Cosmology Centre, Niels Bohr Institute, Copenhagen University, Juliane Maries Vej 30, 2100 Copenhagen O, Denmark\\
$^{2}$Excellence Cluster Universe, Technische Universit\"at M\"unchen,
  Boltzmannstrasse 2, 85748 Garching, Germany\\
$^{3}$Marie Kruses Skole, Stavnsholtvej 29-31, DK-3520 Farum, Denmark}

\begin{document}

%\date{Accepted . Received ; in original form}

%\pagerange{\pageref{firstpage}--\pageref{lastpage}} \pubyear{2010}
\pagerange{1 -- 1} \pubyear{2016}

\maketitle

%\label{firstpage}

\begin{abstract}
The quasar Q0918+1636 ($z$=3.07) has an intervening high-metallicity Damped
Lyman-$\alpha$ Absorber (DLA) along the line of sight, at a redshift of 
$z$=2.58. The DLA is located at a large impact parameter of 16.2 kpc,
and despite this large impact parameter it has a very high metallicity
(consistent with solar). In this paper it is investigated whether a novel
type of galaxy formation models, based on hydrodynamical/gravitational
TreeSPH simulations invoking a new SNII feedback prescription, the 
Haardt \& Madau (2012) ultra-violet background radiation (UVB) field and 
explicit treatment of UVB self-shielding effects, 
can reproduce the observed characteristics of the DLA. 
Effects of UV radiation from young stellar populations in the galaxy, in
particular in the photon energy range 10.36-13.61 eV (relating to
Sulfur II abundance), are also considered in the analysis.
It is found that a) for $L$$\sim$$L_{\star}$ galaxies (at $z$=2.58), about 10\%
of the sight-lines through the galaxies at impact parameter $b$=16.2 kpc
will display a Sulfur II column density of $N(\rm{SII)} \ge 10^{15.82}$ 
cm$^{-2}$ (the observed value for the DLA), and 
b) considering only cases where a near-solar metallicity will be detected at
16.2 kpc impact parameter, the (Bayesian) probability distribution of galaxy
star formation rate (SFR) peaks near the value observed for the
DLA galaxy counterpart of $27^{+20}_{-9} \Msun$/yr. 
It is argued, that the bulk of the $\alpha$-elements, like Sulfur, 
traced by the high metal column density, $b$=16.2 kpc absorption lines, have 
been produced by evolving stars in the inner galaxy, and subsequently 
transported outward by galactic winds.

\end{abstract}


\section{Introduction}

Using optical and near-infrared imaging and spectroscopy,  galaxies at 
redshifts
$z\ga1$ can be detected in two main ways: a) in absorption against the light of
background QSOs, e.g., \cite{1981ARA&A..19...41W,2005ARA&A..43..861W}
and b) in emission, e.g.,
\cite{2002ARA&A..40..579G, 2011ARA&A..49..525S}. However, combining
the information from absorption and emission lines is still a poorly developed
field. Although more than 10\,000 of the so-called Damped Lyman-$\alpha$
Absorbers (DLAs) have been found so far 
\citep{PS04, PW09, 2012A&A...547L...1N}, and
despite some progress \citep[e.g.,][]{2002ApJ...574...51M} in finding their
galaxy counterparts, we still have less than a dozen examples of such
absorption selected galaxies \citep[][see also Rauch et al.\ 2008,
Rauch \& Haehnelt 2011 and Schulze et al.\ 2012]{2012MNRAS.424L...1K}.

In this paper, the main focus will be on the interesting case of a DLA at
$z=2.583$, with a potential galaxy counterpart located such, that the
QSO line-of-sight (los) relative to the galaxy corresponds to a very large 
impact parameter of 16.2 kpc. 
At the same time, the gas abundance of non-refractory elements, such
as Zinc and, to some extent, also Sulfur is close to solar, see 
\citet{2011MNRAS.413.2481F} for details. Further investigations of the 
$z=2.583$ DLA galaxy were carried out by \citet{2013MNRAS.436.361F}. They
performed deep multi-band imaging of the galaxy using HST, NOT/Alfosc and
NOT/NotCam. From SED fitting they derived a star-formation rate (SFR) of
$27^{+20}_{-9}$ $M_{\odot}yr^{-1}$ assuming a \citet{Chabrier2003} initial
mass function (IMF). To search for emission lines from the DLA galaxy, 
VLT/X-Shooter spectroscopy was also performed. A range of emission lines
were clearly detected, with the [\ion{O}{iii}]\,$\lambda$5007 emission line 
having the highest S/N ratio. This line is just $36\pm20$ km s$^{-1}$ 
blue-shifted compared to the center of the low-ionization DLA absorption lines,
very strongly suggesting a physical connection between the DLA absorption
region and the (potential) DLA galaxy. Other emission lines detected (albeit
at somewhat lower S/N) include  the H$\alpha$, H$\beta$, 
[\ion{O}{iii}]\,$\lambda$4960, and [\ion{O}{ii}]\,$\lambda$3727 lines.
 
From a theoretical point of view, it is of great interest to test whether
state-of-the-art galaxy formation models can reproduce, even as an extreme
case, single- or multi-component galaxies displaying such a large metal 
abundance
at 16.2 kpc galactocentric distance. And, if so, it is of interest to
calculate the probability distribution of (semi-)observable quantities,
such as absolute visual magnitude, $\rm{M}_V$, or star-formation rate, SFR,
to check whether the properties observationally inferred for the $z=2.583$
DLA galaxy are in fact ``likely''.   

To address these questions, a large number of new galaxy formation simulations 
have been performed, using a hydrodynamical/gravitational galaxy formation code
with a novel implementation of state-of-the-art feedback models. 
In addition, a previously
performed cosmological simulation has been re-analyzed, to further address
the above issues. The results obtained have been combined with published 
$z \sim 2.25-3$ luminosity functions to calculate DLA galaxy probability
distributions. 

In this paper, details of the simulations and the results 
obtained are presented.  
The paper is organized as follows:
The code and the simulations are described in section 2, and the results
obtained are presented in section 3 and discussed in section 4. In
section 5 it is analyzed how the high metallicity gas in outer galaxy 
was enriched, and, finally, section 6 constitutes the conclusion.

In the appendix, observational models of $z$=2.58 luminosity functions,
as well as procedures of extinction correction are described.

%\section{The code and simulations}
%\label{codesim}

\section{The hydrodynamical/gravitational code and the simulations}
\label{code}

The code used for the simulations is a significantly improved version of
the TreeSPH code we have used previously for galaxy formation simulations 
\citep{SL.03}: Full details will be given in a forthcoming paper; here we
just summarize the main improvements
(1) In lower resolution regions (which will always be
present in cosmological CDM simulations) an improvement in the numerical
accuracy of the integration of the basic equations is obtained by
solving the entropy equation rather than the thermal energy equation ---
we have adopted the ``conservative'' entropy 
equation solving scheme suggested
by \cite{SH02}. 
(2) High-density gas is, subject to a number of conditions, turned into stars 
in a probabilistic way, and in a star-formation event, an SPH particle is 
converted fully into a star particle. The star-formation threshold density is 
set to $n_H = 1$ cm$^{-3}$, and the star-formation efficiency is set to 0.1. 
In conjunction with the new stellar feedback prescription, detailed in 
subsections \ref{fb_implement}-\ref{ESF}, this star-formation
prescription reproduces the observed Kennicutt-Schmidt law \citep{K98}
quite well --- full detail will be given in a forthcoming paper.      
(3) Non-instantaneous recycling of
gas and heavy elements is described through probabilistic ``decay'' of star 
particles back to SPH particles as discussed by \cite{L.02}. 
In a decay event a star particle is converted fully into an SPH particle.
(4) Non-instantaneous chemical evolution tracing
10 elements (H, He, C, N, O, Mg, Si, S, Ca and Fe) has been incorporated
in the code following Lia et~al.\ (2002a,b); the algorithm includes 
supernov\ae\ of type II and type Ia, and mass loss from stars of all masses.
Metal diffusion in Lia et al.\ was 
included with a diffusion coefficient $\kappa$ derived from models of the 
expansion of individual supernova 
remnants. A much more important effect in the present simulations, however,
is the redistribution of metals (and gas) by means of star-burst driven
``galactic super--winds'' (see point 6). This is handled self-consistently by
the code, so we set $\kappa$=0 in the present simulations. 
(5) Atomic radiative cooling depending both on the metal abundance
of the gas and on the meta--galactic UV field (UVB), modeled after \cite{HM12},
is invoked. Radiative transfer (RT) of the UVB is invoked in the way proposed 
by \cite{R13+}. In brief, \cite{R13+} showed that the effect of shielding of 
the UVB by neutral gas can be well described as a function of redshift and the 
local gas density. The effect of UVB shielding on the radiative cooling and
heating functions is explicitly incorporated using CLOUDY \citep{F.13} 
(6) Stellar feedback is incorporated in a new way relative to, e.g., 
\cite{SL.03} and \cite{SL.05}. During the first $\sim$3.4 Myr after a stellar
population has been born, but before the first core collapse supernovae go
off, the massive stars feed radiation energy back to the surrounding ISM.
This affects the star formation rate in the surrounding ISM through 
photo-heating the cold gas, which in turn leads to a decrease in the 
star-formation rate \citep{S.13}. After $\sim$3.4 Myr, the first SNII start
exploding and feeding energy back into the ISM. A classic ``sub-grid''
super-wind model is used to describe the evolution of the surrounding ISM,
receiving large amounts of energy. 

Both aspects of the stellar feedback are described in detail in the following.

\subsection{Stellar feedback}
\label{feedback}

The ``over-cooling catastrophe'' initially plagued simulations of disk galaxy 
formation.  Fully cosmological numerical galaxy simulations consistently 
contained a massive central concentration of stars \citep{Navarro1991}, and
the sizes and angular momenta of the disks were too small 
\citep[e.g.,][]{SL.03}.

Stellar feedback is the favored way of reducing star formation and launching 
outflows \citep{Scannapieco2008, Schaye2010}. Semi-analytic models have found 
that 
significant amounts of stellar feedback are required to match the low mass end 
of 
the luminosity function \citep{Somerville1999, Benson2003, Bower2006, 
Bower2008, 
Bower2012}. \citet{Dutton2009} showed that stellar feedback can remove low 
angular 
momentum material.  

In hydrodynamical simulations, two methods are commonly used to model stellar 
feedback.  One is kinetic feedback that adds velocity kicks to gas particles 
to 
remove them from the inner regions of galaxy disks \citep{SH03, 
Oppenheimer2006, 
DallaVecchia2008}.  The other is thermal feedback in which stars simply heat 
gas 
particles and allow the adiabatic work of the particles to push other gas out 
of
the star forming region
\citep{Gerritsen1997, Thacker2000, Kawata2003, SL.03, S.06}.  

Since stars form in dense regions, the cooling times of the surrounding gas 
are 
short, and without help, the gas will quickly radiate away all the supernova 
energy \citep{katz92}.  In real galaxies, the amount of gas necessary to 
exert 
a dynamical influence on the ISM small.  In simulations, for resolution 
reasons,
such small amounts of gas 
are difficult to model, so a common technique has been to turn off radiative 
cooling for 
a limited amount of time \citep{Gerritsen1997, Thacker2000, SL.03, Brook2004, 
S.06}, but see \cite{H.14} for an alternative approach.

\cite{SL.03} turned off radiative cooling for the $N_n$=50 SPH particles 
nearest to the
star particle providing the feedback. This was a choice in line with 
principles 
of SPH, since all hydrodynamical quantities are smoothed over this number of
SPH particles inside of the SPH smoothing kernel. It is easy to show, that for
gas of hydrogen number density $n_0$ being represented by SPH particles of mass
$m_{\rm{SPH}}$, the radius of the smoothing kernel is given by
\begin{equation} \label{eq:rsmooth}
r =  336 \left(\frac{m_{\rm{SPH}}}{10^5 \Msun}\right)^{1/3} 
\left(\frac{N_n}{50}\right)^{1/3} n_0^{-1/3} ~~ \rm{pc}~~,
\end{equation}
where $N_n$ is the number of SPH particles required to be inside the smoothing 
kernel. When the most massive stars of a newly born stellar population explode
as supernovae type II, the true size of the region affected by the supernova
feedback is considerably less than the size given by eq. (\ref{eq:rsmooth}), as
will be discussed in detail below. This implies that the effect of the 
supernova
feedback on gas near the stellar population is underestimated, due to the SNII
energy being smoothed over the entire smoothing kernel. This turns out to 
hamper
the onset of gas outflows, as discussed by, e.g., \cite{S.06}.

Motivated by these considerations, in this paper we introduce an approach based
on feeding SNII energy back to a region of a size, which is some fixed 
fraction 
of the time-dependent radius of a classic super-wind around a star burst, 
$R_{\rm{SW}}(t)$. This is reasonable, since all core collapse supernovae do not
go off at the same time --- rather the 100 $\Msun$ stars (assumed to be the 
most 
massive stars born in a stellar population) go off after about 3.4 Myr, whereas
9 $\Msun$ stars (assumed to be the least massive stars that explode as SNII), 
go off after about 34 Myr \citep{L.02}. During this 30.6 Myr SNII explosion 
period, the luminosity of the star burst is approximately constant --- see 
further below. 

Moreover, as only part of the gas inside of a blast wave or super-wind will 
remains
hot over a period of about 31 Myr, it makes sense to switch off radiative 
cooling 
inside only a fraction of the classic super-wind radius, as will be discussed 
in the
following. Firstly, blast waves will be quantitatively discussed, as detailed 
numerical simulations are available in this case. This will be followed by a 
more qualitative discussion of the super-wind case.

\subsection{The new supernova feedback prescription - theoretical 
considerations}
\label{fb_theory}

\subsubsection{Blast waves}
\label{bw}
To illustrate the effect of radiative cooling on the shocked region around
a star-burst, it is first (incorrectly) assumed that all SNII energy is 
liberated at the same time, i.e. instantaneously in a small region. 
Furthermore, it is assumed that the ambient gas is homogeneously distributed, 
and that the release of explosion energy is isotropic.
Initially, the radius of outward propagating shock-front is well described by 
the (adiabatic) Sedov-Taylor solution, viz.   
\begin{equation} \label{eq:ST}
R_{\rm{ST}}(t) =  \left(\frac{\xi E}{\rho_0}\right)^{1/5} t^{2/5}~~,
\end{equation}
where $E$ is the total energy released by the explosion, $\rho_0$ is the mass 
density of the surrounding gaseous medium; and $\xi=2.026$ for an adiabatic 
index $\gamma = 5/3$ \citep{OM88}. At about the time $t_{\rm{sf}}$, the shock
becomes radiative, and a cool dense shell is formed behind the shock. An 
expression for this shell formation time is given by \cite{CMB88} (CMB88):
\begin{equation}  \label{eq:tsf}
t_{\rm{sf}} =  3.61 \times 10^4 \frac{E_{51}^{3/14}}{\zeta_m^{5/14}n_0^{4/7}} 
   ~~~~ \rm{yr},
\end{equation}
where $n_0$ is the ambient hydrogen number density in cm$^{-3}$, and $E_{51}$ 
is the explosion energy in units of 10$^{51}$ ergs. The radiative cooling 
function has been roughly approximated by $\Lambda = 1.6 \times 10^{-22} 
\zeta_m T_6^{-1/2}$ ergs cm$^3$ s$^{-1}$, where $\zeta_m = 1$ corresponds to
solar metallicity and $T_6$ is the temperature in units of 10$^6$ K.  

At about $t_{\rm{sf}}$ the shocked region enters a new evolutionary stage that
typically lasts for more than 10$t_{\rm{sf}}$: the pressure-driven snowplow 
(PDS) stage, in which the shell is driven outward by the pressure of the 
interior hot, dilute gaseous bubble - note though that due to radiative cooling
and pressure effects, the mass of hot gas decreases during this stage, as will
be discussed below. For $t_{\rm{sf}} < t < 13t_{\rm{sf}}$, CMB88 find that the
radius of the shocked region is well approximated by an offset power law:
\begin{equation} \label{eq:RS}
R_{\rm{S}}(t) = R_{\rm{PDS}}(\frac{4}{3}t_{\star}-\frac{1}{3})^{3/10} ~~,
\end{equation}
where $R_{\rm{PDS}} = R_{\rm{ST}}(t_{\rm{PDS}} = t_{\rm{sf}}/e$) ($e$ is the 
base of the natural logarithm) and $t_{\star}=t/t_{\rm{PDS}}$. 

Subsequently, the $P dV$ work done by the interior gaseous bubble on the outer,
cool high-density shell becomes insignificant, and the shell expands at a 
slightly slower rate, as it enters the momentum conserving snowplow (MCS) 
phase, with the asymptotic behavior $R_{\rm{S}} \propto t^{1/4}$ 
\citep{O51}, though CMB88 argue that in practice the exponent never reaches
the asymptotic MCS value of 0.25. Finally, when the shock/shell speed becomes 
comparable to the sound speed of the ambient gaseous medium, the ambient
pressure starts to affect the shock propagation, and the expansion
stops. According to \cite{MO77} this time can be expressed as 
\begin{equation} \label{eq:tE} t_{\rm E} =
10^{5.92}E_{\rm 51}^{0.31}n_0^{0.27}\tilde{P}_{\rm 04}^{-0.64} {\rm yr} ~~,
\end{equation} 
where $\tilde{P}_{\rm 04} = 10^{-4}P_0 k^{-1}$, $P_0$ is the ambient pressure 
and k is Boltzmann's constant. CMB88 obtain a similar result for solar 
abundance --- the abundance dependence is in any case very weak.
The region continues to cool radiatively
even after it stops expanding.  \cite{MO77} estimate that the time that
the hot, low density interior will survive is approximately given by
\begin{equation} t_{\rm max} =
10^{6.85}E_{\rm 51}^{0.32}n_0^{0.34}\tilde{P}_{\rm 04}^{-0.70} {\rm yr} ~~.
\end{equation} 
CMB88, and subsequently also \cite{HSL95}, find the for 
$t_{\rm{sf}} < t < 13t_{\rm{sf}}$ the thermal energy of
the hot, dilute gaseous interior decreases with time as
\begin{equation} \label{eq:ehot} 
E_{\rm{th}}(t) =  4.4 \times 10^{50} E_{51} 
    \left(\frac{t}{t_{\rm{sf}}}\right)^{-1.04} ~~ \rm{ergs}.
\end{equation}
\cite{HSL95} showed, using the above expressions from CMB88, that the
mass fraction of hot gas (defined as gas of $T > 10^5~$K), for 
$t_{\rm{sf}} < t < 13t_{\rm{sf}}$ can be expressed as
\begin{equation}\label{eq:epshot} 
\epsilon_{\rm{hot}}(t) =  \frac{T_0}{\bar{T}_{\rm{hot}}} 
    \left(\frac{t}{t_{\rm{sf}}}\right)^{-1.94} n_0^{2/7} 
    \zeta_m^{3/7} E_{51}^{1/7} ~~, 
\end{equation}
where $T_0 = 9.2 \times 10^5$ K, and $\bar{T}_{\rm{hot}}$ is the mean 
temperature of the hot gas. For the energies, densities and metallicities 
considered in this work, $\bar{T}_{\rm{hot}} = 5-10 \times 10^5$ K - see 
\cite{HSL95} for details.  

\begin{figure}
\includegraphics[width=0.48\textwidth]{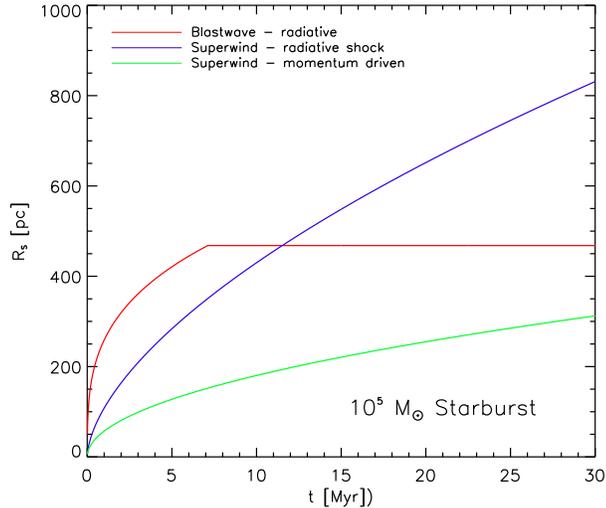}
\caption{Evolution of blast wave and super wind outer shock radii with time,
for a $M_{\star} = 10^5 \Msun$ ``starburst'', and $n_0 = 1$. Red curve
corresponds to a blast wave with effects of radiative cooling included, blue
curve corresponds to a super-wind with a radiative shock and green curve
corresponds to a momentum driven super-wind, such that effects of thermal
pressure are neglected.  
\label{fig:rsh}}
\end{figure}

Due to the strong decrease of hot gas mass fraction with time, it is clear
that only a minor fraction of the hot gas available at the time of shell
formation is still hot when the shell stops expanding (eq. \ref{eq:tE}).
As the hot interior continues cooling radiatively after $t = t_{\rm E}$, an
even smaller fraction of the ``initial'' hot gas survives hot till $t =
t_{\rm{SNII,end}}$ = 30.6 Myr, which for the cases considered here is later
than $t_{\rm E}$. For continuity reasons alone, is is clear that the surviving 
hot gas must be located in the inner parts of the shocked region at $t \sim 
t_{\rm{sf}}$ - this is also the region where the density is very small,
and hence the cooling time very large --- e.g., \cite{LL87}.

To estimate the size of the region that survives as hot gas, we consider the
specific case of a $M_{\star} = 10^5 \Msun$ ``starburst'', exploding in
gas of $n_0 = 1$ cm$^{-3}$, and of solar abundance, i.e., $\zeta_m = 1$.
 
The number of
stars of mass $M \ge 9 \Msun$ (assumed to result in core-collapse supernovae)
that are formed can be expressed as $N_{SNII} = \nu_{100} M_{\star}/(100 \Msun)$,
where $\nu_{100}$ depends on the IMF adopted. For the Chabrier (2003) IMF, as 
implemented in this work, it is found that $\nu_{100} = 1.03$. Assuming that
a total energy of $10^{51}$ ergs is deposited in the ambient gaseous medium per
SNII, this results in $E_{51} = 1030$. From eq. (\ref{eq:tsf}) it follows that
$t_{\rm{sf}} =  1.6 \times 10^5$ yr, and from eq. (\ref{eq:ST}) and the definition
above that $R_{\rm{PDS}} = 102$ pc --- it has been assumed that 
$\rho_0 = 2.3 \times 10^{-24} n_0$ g/cm$^{3}$, which is valid for solar 
abundance. From eq.(\ref{eq:RS}) it follows that 
$R_{\rm{sf}} \equiv R_{\rm{S}}(t_{\rm{sf}})=
146$ pc, and the region will subsequently keep expanding until $t \simeq 
t_{\rm E} = 7.1$ Myr ($\tilde{P}_{\rm 04} = 1$ is adopted).
  
The analytical model of CMB88 indicates that their results can be extended to
even later times than specified above, i.e., $t > 13t_{\rm{sf}}$. Hence, for the purposes of the order of magnitude estimates given in this work, we assume that
eq.(\ref{eq:RS}) describes the evolution of the blast wave/shell till 
$t = t_{\rm{E}} = 7.1$ Myr ($\simeq 44t_{\rm{sf}}$).  It follows that 
$R_{\rm{E}} \equiv R_{\rm{S}}(t_{\rm{E}}) = 468$ pc --- this will be somewhat 
of an overestimate,
since the effect of the pressure of the ambient gas will be to reduce the
expansion speed at $t \sim t_{\rm{E}}$ - we comment on this below. It is, for
simplicity, moreover assumed that $R_{\rm{S}}(t>t_{\rm{E}}) = 
R_{\rm{E}}$ until $t = t_{\rm{SNII,end}}$ (in reality the shell
radius will initially oscillate around the value given above, and eventually,
as radiative cooling removes the pressure support from the interior, hot gas,
start to implode at $t \sim t_{\rm max} = 65$ Myr --- e.g., \cite{MO77}).
The evolution of the blast wave shock radius with time, under the assumptions
specified above, is shown in Fig.~\ref{fig:rsh}. 
 
At $t = t_{\rm{E}}$, the mass of the blast wave region will be 
$M_{bw}(t_{\rm{E}}) \simeq
\frac{4 \pi}{3} R_{\rm{E}}^3 \rho_0 = 1.5 \times 10^7 \Msun$. This is somewhat
of an overestimate, since $R_{\rm{E}}$ is overestimated --- see above. 
Using eq.(\ref{eq:epshot}), it follows that an upper limit to the mass of hot 
gas at $t = t_{\rm{E}}$ is 
$M_{\rm{hot}} \simeq \epsilon_{\rm{hot}}(t_{\rm{E}})$ $M_{bw}(t_{\rm{E}})
= 3.4 \times 10^4 \Msun$ (a value of $ \bar{T}_{\rm{hot}} = 6.8 \times 10^5$ K
has been adopted --- this follows from the results and scaling relations given
in \cite{HSL95}). At the time of shell formation, $t_{\rm{sf}}$,
the mass of the blast wave region is $M_{bw}(t_{\rm{sf}}) \simeq 
\frac{4 \pi}{3} R_{\rm{sf}}^3 \rho_0 = 4.4 \times 10^5 \Msun$. It follows
that an upper limit to the fraction of gas at $t = t_{\rm{sf}}$, that is 
still hot at $t = t_{\rm{E}}$, is about $M_{\rm{hot}}/M_{bw}(t_{\rm{sf}}) =
0.077$. Moreover, at times later than $t_{\rm{E}}$ the thermal energy will
continue decreasing --- see above. Assuming, for lack of more detailed 
information, that the subsequent evolution of the thermal energy is described 
by $E_{\rm{th}}(t) \propto t^{-1.04}$, (cf., eq.(\ref{eq:ehot})), and that
the mean temperature of the hot gas is approximately constant, it follows 
that, at $t = t_{\rm{SNII,end}}$, $M_{\rm{hot}} \simeq 7.4 \times 10^3 \Msun
\simeq 0.017 M_{bw}(t_{\rm{sf}})$. So, only a few percent of the gas 
available at the time of onset of shell formation remains hot till $t = 
t_{\rm{SNII,end}}$. Approximating the gas distribution at $t = t_{\rm{sf}}$
by the Sedov-Taylor profile for adiabatic index $\gamma = 5/3$ (see, eg.,
\cite{LL87} for details), it can be shown that only the gas situated inside of
0.6 $R_{\rm{sf}}$ remains hot till $t = t_{\rm{SNII,end}}$.

So, as a consequence of the arguments given above, it seems reasonable to
only switch radiative cooling off for SPH particles located inside some
fraction (denoted $\beta$ in the following) of $R_{\rm{S}}(t)$.
  
\subsubsection{Star-burst driven super winds}
\label{sw}
In reality, simultaneously formed massive stars of different masses do not
explode as SNIIs at the same time. The most massive O stars considered in
this work, of mass 100 $\Msun$, explode after about $t_{60} = 3.4$ Myr, whereas
the least massive stars that explode as SNIIs, assumed here to be B stars of 
mass 9 $\Msun$, explode after about $t_{8} = 34$ Myr. For standard IMFs 
(note that the \cite{Chabrier2003} and \cite{Salp55} IMFs have very similar
slopes at the high mass end), the resulting SNII luminosity is approximately
constant during the time interval $t_{60} \le t \le t_8$ (see, e.g.,
\cite{MK87}), so for simplicity, the luminosity will be assumed to be 
constant in the following.

For an adiabatic super wind, the evolution of the shock radius is given by
\begin{equation} \label{eq:AD}
R_{\rm{SW}}(t) =  \alpha \left(\frac{L_{\rm{SW}}}{\rho_0}\right)^{1/5} t^{3/5}~~,
\end{equation}
where $L_{\rm{SW}} = \dot{E}_{SNII}$ is the luminosity of the star burst, and 
$\alpha = 0.88$ \citep[e.g.,][]{W.77}. After a relatively 
short time, the shock becomes radiative. This time can be expressed as
\begin{equation} \label{eq:tcool}
t_{\rm{cool}} =  3.4 \times 10^4 \left(\frac{L_{\rm{SW},39}}{n_0}\right)^{1/2}~~ 
\rm{yr}~~,
\end{equation}
where $L_{\rm{SW},39}$ is the luminosity in units of 10$^{39}$ ergs/sec 
\citep{C.75}
Subsequently, as long as radiative cooling of the hot interior is not
important, the evolution of the shock radius is given approximately by
\begin{equation} \label{eq:SW}
R_{\rm{SW}}(t) =  \alpha \left(\frac{L_{\rm{SW}}}{\rho_0}\right)^{1/5} t^{3/5} =
105 \left(\frac{L_{\rm{SW},39}}{n_0}\right)^{1/5} t_6^{3/5}~~\rm{kpc} ~~,
\end{equation}
where $\alpha = 0.76$ has been assumed \citep[e.g.,][]{W.77}, and $t_6$ is
the time in Myrs. Given that the adiabatic phase is so brief, eq.(\ref{eq:SW})
will be used in the following to express the evolution of the super-wind shock 
radius. During the radiative shock phase, the thermal energy of the super-wind 
region is given by $\frac{5}{11} L_{SW} t$, i.e., at any time about 45$\%$ of
the total amount of super-nova energy released.

At some point in time, radiative cooling of the interior will start to affect
the expansion of the super-wind. Expressions for the temporal evolution of
the thermal energy, like eq. (\ref{eq:ehot}) and (\ref{eq:epshot}) above, are
unfortunately not available in the literature. However, one can estimate the
evolution of the shock radius with time, in the case of complete radiative 
cooling, as follows: 

In general, under simplifying conditions, the time derivative of
total (radially outward directed) momentum of the super-wind region can be
expressed as   
\begin{equation} \label{eq:motion}
\frac{d}{dt} M \bar{v} \simeq  4 \pi R_{\rm{SW}}^2 \bar{p} + \dot{M}_{\rm{SB}} 
v_{\rm{SB}} ~~,
\end{equation}
where M is the total mass of the super-wind region, $\bar{v}$ is the mass
averaged radial velocity, $\bar{p}$ is the volume averaged pressure, 
$\dot{M}_{\rm{SB}}$ is the rate of mass ejection by the central star-burst and
$v_{\rm{SB}}$ is the average mass ejection velocity, assumed to be constant.
$\dot{M}_{\rm{SB}} v_{\rm{SB}}$ is the momentum injection rate in the central
star-burst region. The main assumptions made are 1) that the shock is strong,
and 2) that the region of momentum injection is much smaller than 
$R_{\rm{SW}}$. In the case where the average pressure can be neglected, due
to effects of radiative cooling, eq. (\ref{eq:motion}) simplifies to   
\begin{equation} \label{eq:momentum}
\frac{d}{dt} M \bar{v} \simeq \dot{M}_{\rm{SB}} v_{\rm{SB}} = 
\frac{2 \eta L_{\rm{SW}}}{v_{\rm{SB}}} ~~,
\end{equation}
where $\eta$ is fraction of total star-burst energy, that comes out in the 
form of kinetic energy. For the Sedov-Taylor solution, $\eta \simeq 0.28$.
As most of the swept-up mass is concentrated in a thin shell behind the
outer shock, $\bar{v} \simeq v_{\rm{SW}}$, where $v_{\rm{SW}}$ is the shock
speed. Hence, eq. (\ref{eq:momentum}) can be integrated to yield
\begin{equation} \label{eq:MO}
R_{\rm{SW,kin}}(t) \simeq  \left(\frac{3 \eta L_{\rm{SW}}}{\pi \rho_0 v_{\rm{SB}}}\right)^{1/4} t^{1/2}~~.
\end{equation}
Inserting characteristic values yields
\begin{equation} \label{eq:MOC}
R_{\rm{SW,kin}}(t) \simeq  45 \left(\frac{\eta_{0.28} L_{\rm{SW},39}}
{n_0 v_{\rm{SB},3000}}\right)^{1/4} t_6^{1/2}~~\rm{pc}~,
\end{equation}
where $\eta_{0.28} = \eta/0.28$ and $v_{\rm{SB},3000} = v_{\rm{SB}}/3000$ km/s.
For typical values, $L_{\rm{SW},39} \sim 1$ (see below) and  
$v_{\rm{SB},3000} \sim 1$ (from supernova energetics), 
$R_{\rm{SW,kin}}/R_{\rm{SW}} \la 0.5$ for $t \ga 0.2$ Myr. 

To be specific, we again consider the
case of a $M_{\star} = 10^5 \Msun$ star-burst, located in ambient
gas of $n_0 = 1$ cm$^{-3}$. Assuming constant luminosity, the luminosity
will be given by
\begin{equation} \label{eq:lumSB}
L_{\rm{SW}} =  \left(\frac{\nu_{100} M_{\star} E_{\rm{SN}}}{(100 \Msun) t_{\rm{SNII,end}}}\right) =
1.07 \times 10^{39} \left(\frac{M_{\star}}{10^5 \Msun}\right) \rm{erg/sec} ~~,
\end{equation}
where $ E_{\rm{SN}}$ is the thermal plus kinetic energy release by one 
supernova, again assumed to be 10$^{51}$ ergs. Hence, $L_{\rm{SW},39}$ = 1.07.
Inserting this, and $n_0 = 1$ in eq.(\ref{eq:SW}), results in the blue
curve shown in Fig.~\ref{fig:rsh}. When the shock speed eventually becomes 
comparable to the sound speed of the ambient gaseous medium, the ambient
pressure starts to affect the shock propagation, halting the expansion.
As can be seen from the figure, this happens at a somewhat later time, 
than for the blast wave (comparing the slope of the red curve at 
$t = t_{\rm{E}}$ to the slope of the blue curve at any $t$), so, 
for simplicity, this has not been incorporated
in the figure. Also shown in the figure, by the green curve, is the 
momentum-driven case, where the effect of the pressure of the interior is 
neglected. The shock speed becomes comparable to the sound speed after a 
few Myr, so at later times, the green curve is a conservative upper limit
to the actual radius of the super-wind region for this (albeit extreme) case.

Lacking detailed numerical simulations, we can only state that it quite
likely that the actual radius of the super-wind region, $R_{\rm{SW}}(t)$, is 
somewhat less than the expression given above by eq.(\ref{eq:SW}). The radius,
within which the hot gas does not cool radiatively during the entire supernova 
feedback period, $t_{\rm{SNII,end}}$, is obviously even less than this.
So, as for the blast wave case, it seems reasonable to
only switch radiative cooling off for SPH particles located inside some
fraction, $\beta$, of $R_{\rm{SW}}(t)$ (eq. \ref{eq:SW}).
    
\subsubsection{Super wind breakthrough and breakout from the disk}
\label{break}
At present, the thin, dense part of the Galactic disk has a scale height of
about 200 pc \citep[e.g.,][]{H90}. When the radius of the super wind becomes
comparable to this scale height, the super wind will start moving in directions
perpendicular to the disk \citep[e.g.,][]{KM92, MF99}, and ``breakthrough'' 
has occurred.
According to Koo \& McKee (1992) this occurs if the star-burst results in at
least $\sim$50 SNII's. If the burst results in at least $\sim$1000 SNII's,
the vertically moving super wind will move through the much more extended
(scale height $\sim$1 kpc), but lower density ($n_0 \sim 0.05$ cm$^{-3}$) HII 
layer as well, and complete ``breakout'' occurs, in which significant amounts
of shocked gas completely leaves the galaxy \citep{KM92}.

\subsection{The new supernova feedback prescription - implementation}
\label{fb_implement}
As discussed in sections \ref{bw}-\ref{break}, 
a $M_{\star} = 10^5 \Msun$ star burst, results in about 
1000 SNII's, and hence super wind breakout is likely to occur. As shown by
the above authors, when the super wind radius reaches about the scale height
of the disk, the geometry of the super wind will change from spherical to 
approximately that of a bi-polar outflow, perpendicular to the plane of the 
disk. As can be seen from Fig.~\ref{fig:rsh}, $R_{\rm{SW}} \sim$ 200 pc
already at $t \sim 3$ Myr for the radiative shock super wind, in cases where a
cold gaseous disk has been established. Hence, after
that time it would be incorrect to switch radiative cooling off for all
SPH particles inside of $R_{\rm{SW}}(t)$, as some of these particles will
never be affected by the super wind. As can be seen from the Figure, at
$t \sim t_{\rm{SNII,end}}$ $R_{\rm{SW}} \sim$ 800 pc for the spherically 
symmetric radiative shock super wind. Motivated by this, and the arguments 
given above, radiative
cooling is, at any given time $t$,  only switched off for SPH particles 
located inside of $\beta R_{\rm{SW}}(t)$ (eq. \ref{eq:SW}), where $\beta = 
0.25$ is adopted.

The energy and momentum feedback from the star burst is of course deposited in
the small region containing the newborn massive stars. This region will in 
general be considerably smaller than the scale height of the disk, so one
should in principle, at least at late times, deposit the energy to even fewer 
SPH particles than the ones for which radiative cooling is switched off. 
Following \cite{S.06}, however, we choose to feed star burst energy to all 
particles, for which cooling has been turned off --- extensive testing 
indicates that it makes no significant difference to the results. We 
furthermore require that at least
$N_{\rm{SPH,min}} = 2$ SPH particles have radiative cooling switched off and
receive SNII energy feedback --- variations of $N_{\rm{SPH,min}}$ by a factor
of two makes no significant difference to the results either. We note that,
since the SPH implementation is based on solving the entropy equation, even
injecting all energy into one particle only, still gives approximately the
correct behavior of an explosion --- see \cite{SH02} for details.

\subsection{Early stellar feedback}
\label{ESF}
Thermal supernova feedback alone, even when implemented as described above,
is not sufficient to completely prevent ``over-cooling'', at least when
it is assumed that each SNII releases at most 10$^{51}$ erg. This results
in galaxies with too massive central concentrations of stars, leading in turn
to too centrally peaked rotation curves \citep[e.g.,][]{Stinson2010,
Scannapieco2012}.

Moreover, recently, abundance matching techniques \citep{Conroy2006} have been
used to compare total halo masses with galaxy stellar masses \citep{Conroy2009,
Moster2010,Guo2010,Behroozi2010,Moster2013}. \citet{Guo2010} and 
\citet{Sawala2011} showed that most galaxy formation simulations 
form too many stars compared to what abundance matching predicts. 

Even without over-cooling, \citet{vandenBosch2002} showed that the amount of 
low angular momentum material in collapsed collision-less halos exceeds the 
amount of low angular momentum material observed in disk galaxies.  
Consequently, this low angular momentum material needs to be removed from the 
center of the system.  

During the first 3.4 Myr after a population of stars has been formed, the 
massive stars emit large amounts of (primarily UV) radiation, but no SNII
explosions are taking place. This early stellar feedback (ESF) can modify the 
ISM around
the stellar population, effectively suppressing star formation in a region
around the new born stars. Indeed, \citet{Murray2010} showed that there can be 
significant feedback effects from stars before they explode as supernovae.  
Moreover, \citet{Hopkins2011} incorporated a 
kinetic radiation pressure feedback into simulations of isolated disk galaxies 
and found that the feedback could strongly regulate star formation.  

Motivated by this, \cite{S.13} implemented a scheme based on thermal pressure 
to provide feedback during the time between when stars are formed and the 
first SNs explode --- see also \cite{W.15}.
Using this early stellar feedback (ESF) prescription, 
\citet{Maccio2012} showed 
that the feedback removes low angular momentum dark matter in galaxies up to 
nearly L$^\star$, producing cored dark matter density profiles.  
\citet{Stinson2012} showed that the metal rich outflows created by this 
feedback match observations of OVI in the circum-galactic medium of star 
forming galaxies. Brook et al. (2012a,b) also showed that
a sample of galaxies, of masses less than or similar to that of the Milky Way, 
form disks that follow a wide range of disk scaling relationships.

The radiation feedback from massive stars has been implemented in the code
following \cite{S.13}: To model the luminosity of stars, a simple fit of the 
mass-luminosity relationship observed in binary 
star systems by \citet{torres10} is used:
\begin{equation}
\frac{L}{L_\odot} = 
\begin{cases}
(\frac{M}{M_\odot})^{4},  & M < 10 M_\odot\, \\
100(\frac{M}{M_\odot})^{2},   & M > 10 M_\odot\, \\
\end{cases}
\end{equation}
This relationship leads to about $3\times10^{50}$ ergs of energy being 
released from the high mass stars per M$_\odot$ of the entire stellar 
population over the 3.4 Myr between the formation of the stellar population 
and the commencing of SNII. These photons do not couple 
efficiently with the surrounding ISM \citep{freyer06}, so only a fraction,
$\epsilon_{\rm{ESF}}$ (of the order 10\% --- see further below),
of the radiation energy is injected as thermal energy in the  
surrounding gas. Moreover, radiative cooling is not turned off for this form 
of energy input. 
It is well established that such thermal energy injection is highly 
inefficient at the spatial and temporal resolution of the type of 
cosmological simulations used here \citep{katz92,kay02}. Though the dynamical
effect is limited, ESF effectively halts star formation in the region
immediately surrounding a recently formed stellar population by increasing
the temperature above the threshold temperature for star formation. 
We note that \cite{H.14} also find that the non-linear interaction of
ESF and supernova feedback is critical to explain large-scale outflows,
self-regulation of star formation etc.

\subsection{The simulations}
\label{subsec:sims}

The basis for the hydro/gravity simulations was a cosmological, dark matter 
(DM) only simulation of a ``field-galaxy'' region. 
The cosmological initial conditions were based on a $\Lambda$CDM model with
($\Omega_M,\Omega_{\Lambda}$)=(0.3,0.7) and a Hubble parameter
$H_0 = 100h\,\kmsMpc = 65\,\kmsMpc$, close to the parameters suggested by
the most recent Planck 2015 results,
The simulation had co-moving box length 10~$h^{-1}$Mpc, and
was evolved from a starting 
redshift of $z_i = 39$ to $z = 0$ assuming periodic boundary conditions -- 
more detail on the DM only simulation is given in \cite{SL.03}.
At $z = 0$, the 20 most massive haloes (see further below) were selected --
they span a range of virial masses of $M_\mathrm{vir}
\sim 6\times10^{11} -  4\times10^{12} \Msun$. 
Mass and force resolution was increased in Lagrangian regions enclosing the 
haloes, and in these regions all DM particles were split into a DM particle
and a gas (SPH) particle according to an adopted universal baryon fraction of
$f_b$=0.15, in line with recent estimates. The evolution of these regions was 
then re-simulated using hydro/gravity code described
in section ~\ref{code} -- this approach has been dubbed the ``zoom-in'' 
technique -- full detail is given in \cite{SL.03}. Particle masses were
$m_{\rm{SPH}}$=$m_*$=
7.3x10$^5$ and $m_{\rm{DM}}$=4.2x10$^6$ $h^{-1}$M$_{\odot}$, and gravity spline
softening lengths (approximately force resolution) were 
$\epsilon_{\rm{SPH}}$=$\epsilon_*$
=380 and $\epsilon_{\rm{DM}}$=680 $h^{-1}$pc. The number of
SPH+DM particles per re-simulated halo region lies in the range
2.5x10$^5$-1.5x10$^6$ for these medium-resolution simulations. The ESF
efficiency was set to $\epsilon_{\rm{ESF}}$=0.10 for these runs --- see below. 

The simulations were evolved from a starting redshift of $z_i = 39$ to 
$z = 2.58$. At $z = 2.58$ the 20 simulations contain a total of 158 well
resolved primary haloes (i.e., not counting sub-haloes) of virial masses
$M_\mathrm{vir} \sim 10^{10} -  10^{12} \Msun$. The stellar masses of the
corresponding primary galaxies span the range $M_* \sim 10^7 -  
10^{10} \Msun$.   

In addition to the above 20 re-simulations, another 20 re-simulations
of scaled-up versions of the regions were undertaken. Linear scales and
velocities were scaled by a factor 1.6, masses were correspondingly
scaled by a factor 1.6$^3 \simeq 4.1$. 
Since the $\Lambda$CDM power-spectrum is fairly constant over this limited 
mass range, the rescaling is a reasonable approximation. The reason why
such more massive haloes are also re-simulated is that massive galaxies
are crucial for the purposes of this paper, as will be detailed in the
following. Particle masses were $m_{\rm{SPH}}$=$m_*$=
3.0x10$^6$ and $m_{\rm{DM}}$=1.7x10$^7$ $h^{-1}$M$_{\odot}$, and gravity spline
softening lengths (approximately force resolution) were 
$\epsilon_{\rm{SPH}}$=$\epsilon_*$=610 and 
$\epsilon_{\rm{DM}}$=1090 $h^{-1}$pc. The ESF efficiency was set to 
$\epsilon_{\rm{ESF}}$=0.18 for these runs --- see below.  
At $z = 2.58$ the 20 simulations of the scaled regions contain a total of 20 
well resolved primary haloes of virial masses
$M_\mathrm{vir} \sim 10^{12} -  4$x$10^{12} \Msun$ --- only haloes of 
$M_\mathrm{vir} \ge 10^{12} \Msun$ are considered in the analysis of the
scaled regions.
The stellar masses of the corresponding primary galaxies span the range 
$M_* \sim 10^{10} -  10^{11} \Msun$. 

\begin{figure}
\includegraphics[width=0.48\textwidth]{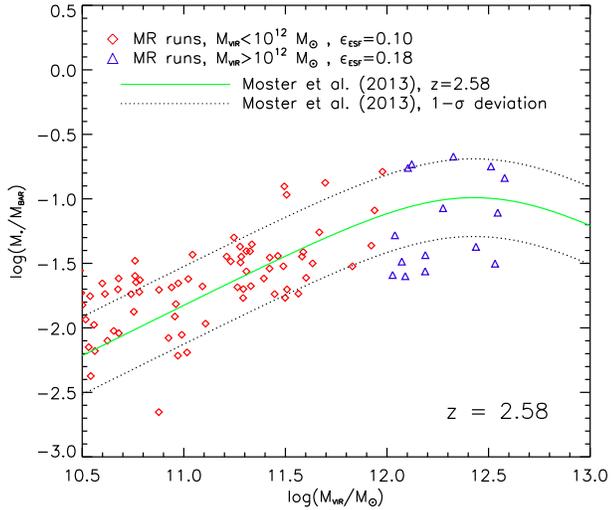}
\caption{$M_*/M_{bar}$ vs. $M_{vir}$ relation at $z$=2.58 for the medium
resolution runs. Most galaxies have been run with $\epsilon_{\rm{ESF}}$=0.10,
but galaxies of $M_{vir}>10^{12} \Msun$ have been run with   
$\epsilon_{\rm{ESF}}$=0.18. Also shown is the \citet{Moster2013} relation
for $z$=2.58, together with the 1-$\sigma$ deviation curves, also from
\citet{Moster2013}.
\label{fig:shm}}
\end{figure}

The ESF efficiencies were chosen such that, at $z$=2.58, the ratio between 
the galaxy
stellar mass and the total baryonic mass available inside of the virial
radius, $M_*/M_{bar}$, as a function of virial mass approximately follows
the relation suggested by \cite{Moster2013} (for $z$=2.58). For the 
scaled-up runs, a few galaxies of $M_*/M_{bar} < 0.02$ were omitted from
the analysis --- this does not affect the analysis in any way. The
resulting $M_*/M_{bar}$ vs. $M_{vir}$ relation at $z$=2.58 is shown in 
Fig.~\ref{fig:shm}. 

Subsequently, for the actual production runs, the 40 simulations were
redone at four times higher mass resolution and 1.6 times higher force 
resolution. For the unscaled haloes, these HR
simulations had particle masses of $m_{\rm{SPH}}$=$m_*$=1.8x10$^5$ and 
$m_{\rm{DM}}$=1.1x10$^6$ $h^{-1}$M$_{\odot}$, and gravity spline
softening lengths were $\epsilon_{\rm{SPH}}$=$\epsilon_*$
=240 and $\epsilon_{\rm{DM}}$=430 $h^{-1}$pc. For the scaled-up versions,
the corresponding values are $m_{\rm{SPH}}$=$m_*$=7.4x10$^5$ and 
$m_{\rm{DM}}$=4.5x10$^6$ $h^{-1}$M$_{\odot}$, $\epsilon_{\rm{SPH}}$=
$\epsilon_*$=380 and $\epsilon_{\rm{DM}}$=680 $h^{-1}$pc.

The number of SPH+DM particles per re-simulated HR halo region was about
1-6x10$^6$. As found by \cite{S.13}, when the numerical resolution is 
increased, the ESF efficiency typically has to be decreased slightly,
to result in a HR stellar galaxy of the same mass as the corresponding MR 
stellar galaxy.
For the HR simulations of $M_\mathrm{vir} \le 10^{12} \Msun$ haloes,
values of $\epsilon_{\rm{ESF}}$=0.08-0.09 were used. For the   
$M_\mathrm{vir} > 10^{12} \Msun$ haloes, $\epsilon_{\rm{ESF}}$=0.15-0.16. 

In order to increase the number of lower mass galaxies, data for 27 HR
simulations, aimed at producing disk galaxies at $z$=0 of stellar masses
comparable to that of the Milky Way or less, were included in the analysis.
The simulations are characterized by $m_{\rm{SPH}}$=$m_*$=0.9-2.8x10$^5$ and 
$m_{\rm{DM}}$=0.5-1.6x10$^6$ $h^{-1}$M$_{\odot}$, and gravity spline
softening lengths were $\epsilon_{\rm{SPH}}$=$\epsilon_*$
=190-280 and $\epsilon_{\rm{DM}}$=340-490 $h^{-1}$pc. For these simulations,
SPH+DM particle numbers are 1-3x10$^6$. At $z = 2.58$ the 27 simulations 
contain a total of 65 well resolved primary haloes of virial masses
$M_\mathrm{vir} \sim $0.3-6x10$^{11} \Msun$. The stellar masses of the
corresponding primary galaxies span the range $M_* \sim$ 10$^8$ -  
3x10$^9 \Msun$. 

It turns out, that the results obtained in this paper are quite insensitive
to whether this additional sample of HR disk galaxy simulations is included or 
not. However, to improve the statistics and the resolution at lower galaxy 
masses, we chose to include it in the following. 

\section{Results and analysis}
\label{sec:results}

\subsection{SII column densities}
\label{subsec:SII}
At $z$=2.58, all galaxies located in the 2 x 20 haloes simulated, as well
as in the HR disk galaxy simulations, are analyzed, subject to a SFR selection
criterion: only galaxies of SFR above a threshold SFR$_{detect}$ are considered
as being potential DLA galaxies --- this is in order to mimic the 
observational situation. For most purposes, we adopt 
log(SFR$_{detect}$) = -0.5 (where the SFR is expressed in $\Msun$/yr), but
we shall also consider other values of SFR$_{detect}$. It turns out,
that the main results obtained are quite insensitive to the choice of this
parameter. As will be shown below, log(SFR$_{detect}$) = -0.5 corresponds to
$M_V \simeq -18.5$, 4-5 magnitudes fainter than (the observational) 
M$_{V,\star}$ at such redshifts. With the above selection criterion, a total
of 95 galaxies are selected at $z$=2.58. 

We shall in this work focus on calculating the column density of singly
ionized Sulfur. The reason for this is threefold: a) Sulfur is only weakly
depleted on to dust grains (e.g., \cite{FS97}, but see also \cite{C.09},
and \cite{J09}, b) the abundance of S is tracked in the simulations, 
and c) N(SII) is determined very accurately for the DLA:
\citet{2011MNRAS.413.2481F} obtain log(N(SII)) = 15.82$\pm$0.01.

For each galaxy selected, 600 sight-lines were shot through the galaxy at
random positions and directions, though constrained such that, for every 
sight-line, the impact parameter was 16.2 kpc. Along each sight-line, the
column density of Sulfur II, $N$(SII), is calculated over a 1000 kpc path,
centered at the galaxy, viz.
\begin{equation}\label{eq:NSIIv1}
N(\rm{SII}) = 
\int_{-l_{\rm{max}}}^{l_{\rm{max}}} \it{n}_{\rm{SII}}(l) ~dl ~~, 
\end{equation}
where $n_{\rm{SII}}(l)$ is the number density of Sulfur II as a function
of the position along the line-of-sight, and $l_{\rm{max}} = 500$ kpc (below
we discuss the contribution from non-local SII). At a given position, the 
fraction of singly ionized Sulfur atoms, $f_{\rm{SII}}$, depends on the
local hydrogen density, $n_H$, temperature, $T$, radiation intensity, 
$J(\nu)$ (where $\nu$ denotes frequency), and also (though weakly), on the 
gas metal abundance. This allows eq. (\ref{eq:NSIIv1}) to be rewritten as
\begin{equation}\label{eq:NSIIv2}
N(\rm{SII}) = 
\int_{-l_{\rm{max}}}^{l_{\rm{max}}} \it{n}_{\rm{S}}(\vec{x})~ 
f_{\rm{SII}}(n_H(\vec{x}),T(\vec{x});J(\nu,\vec{x})) ~dl ~~, 
\end{equation}
where $\vec{x}$ denotes the position corresponding to line-of-sight length $l$,
and the weak dependence on metal abundance has been neglected. Adopting the
mean field approximation of \cite{R13+} it is assumed that the local radiation
field depends mainly on the local hydrogen number density $n_H(\vec{x})$. The
radiation field is modeled as a superposition of a transmitted UVB part and
a (hydrogen) recombination radiation part. At a given redshift z, it is 
assumed that a fraction, $f_{\rm{RR}}$, of the local photo-ionization rate is 
due to recombination radiation and that the remaining fraction of the  
photo-ionization, $f_{\rm{TR}}=(1.0-f_{\rm{RR}})$, is caused by transmitted UBV
radiation. The fractions $f_{\rm{RR}}$ and $f_{\rm{TR}}$ are assumed to
be functions of $n_H$ and $z$, and are modeled in accordance with the results
of \cite{R13+} --- generally, at low $n_H$, transmitted UVB dominates, and, at
large $n_H$ , recombination radiation dominates. 
Finally, based on $n_H$, $T$ and the local radiation field, 
$f_{\rm{SII}}$ is determined at the position of each SPH particle using CLOUDY 
\citep{F.13} --- more detail will be given in a forthcoming paper.

In practice, eq.~(\ref{eq:NSIIv2}) is discretized as
\begin{equation}\label{eq:NSIIv3}
N(\rm{SII}) \approx
\sum\limits_{i=-N_{\rm{max}}}^{N_{\rm{max}}} \it{n}_{\rm{S}}(\vec{x_i})~ 
f_{\rm{SII}}(n_H(\vec{x_i}),T(\vec{x_i});J(\nu,\vec{x_i})) ~\Delta l ~~, 
\end{equation}
where $N_{\rm{max}}$=5000 was adopted, implying 
$\Delta l = l_{\rm{max}}/N_{\rm{max}} = 0.1$~kpc --- this choice of $\Delta l$
was found to give stable results in the discretisation. At each point in space,
$\vec{x_i}$, the number density of Sulfur II is estimated using the
cubic spline smoothing kernel of \cite{ML85}, as
\begin{multline}
n_{\rm{SII}}(\vec{x_i}) = \frac{\rho_{\rm{SII}}(\vec{x_i})}{m_{\rm{S}}} = \\  
\frac{1}{m_{\rm{S}}}
\sum\limits_{j=1}^{N_n} m_j~Z_{\rm{S}}(\vec{x_j})~ 
f_{\rm{SII}}(n_H(\vec{x_j}),T(\vec{x_j});J(\nu,\vec{x_j})) 
W(x_{ij},h_i) ~~, 
\end{multline}
where $m_{\rm{S}}$ is the mass of a Sulfur atom, $m_j$ is the mass of SPH
particle $j$ (all SPH particles have the same mass in the present study), 
$Z_{\rm{S}}(\vec{x_j})$ is the Sulfur abundance (by mass) of SPH particle 
$j$, $x_{ij}=|\vec{x_j}-\vec{x_i}|$, $W$ is the smoothing kernel, $N_n$ is the
number of SPH neighbors in the SPH formalism ($N_n$=50 in the present
simulations), and
$h_i$ is chosen such that if the SPH particles are ordered after increasing 
distance to $\vec{x_i}$, then $2 h_i$ corresponds to the distance halfway
between particle $N_n$ and particle $N_n + 1$. Hence, eq.~(\ref{eq:NSIIv3})
is finally expressed as
\begin{multline}\label{eq:NSIIv4}
N(\rm{SII}) \approx \frac{1}{m_{\rm{S}}} \times 
\sum\limits_{i=-N_{\rm{max}}}^{N_{\rm{max}}}[ \\
\sum\limits_{j=1}^{N_n} m_j~
Z_{\rm{S}}(\vec{x_j})~ f_{\rm{SII}}(n_H(\vec{x_j}),T(\vec{x_j});J(\nu,\vec{x_j})) W(x_{ij},h_i)] ~\Delta l ~.
\end{multline}
A sight-line through a given galaxy (1) at impact parameter $b$ = 16.2 kpc is 
counted if there is not another galaxy (2), of SFR$_2$ $>$ SFR$_{detect}$ and
SFR$_2 \ge $SFR$_1$, situated
such that the impact parameter of the sight-line is less than 16.2 kpc relative
to that galaxy (if the impact parameter of the second galaxy is less than
3 kpc it is assumed that the second galaxy is undetected due to the brightness
of the QSO, so the sight-line is counted, unless a third galaxy full fills the
conditions above).  
 
\begin{figure}
\includegraphics[width=0.48\textwidth]{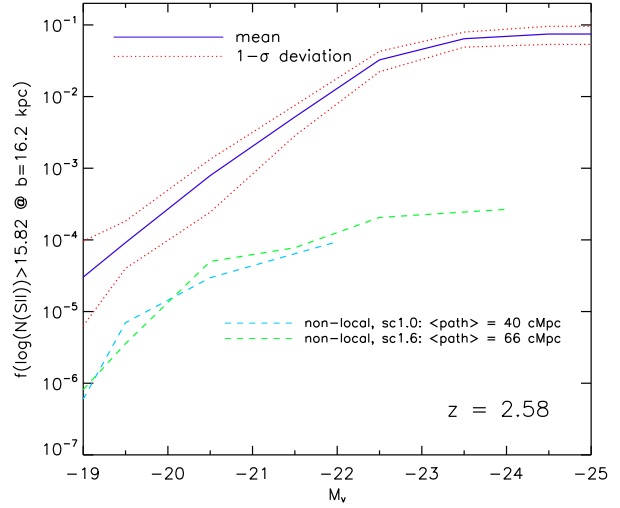}
\caption{Probability of finding log(N(SII))$\ge$15.82 for a sight-line at 
impact parameter $b$ = 16.2 kpc relative to a given galaxy of absolute
magnitude $M_V$ at $z$=2.58. The mean value is shown by the blue line, and
the 1-$\sigma$ (statistical) range is indicated by the dotted red lines.  
Also shown, by the magenta and green dashed lines, are the probabilities
that the SII absorption is caused by non-local galaxies of physical distances
of $0.2<d<11$ Mpc and $0.2<d<18$ Mpc, respectively. This corresponds to
velocity shifts of up to 3000 and 4800 km/s, respectively, for a pure
Hubble flow --- see subsection \ref{subsec:nonloc} for more detail. 
\label{fig:probs}}
\end{figure}

For each galaxy, the fraction of the number of sight-lines with 
log(N(SII))$\ge$15.82 relative to the total number of ``permitted'' (see above)
sight-lines, $p_{15.82}$ is determined. $p_{15.82}$ can be interpreted as the 
probability of finding log(N(SII))$\ge$15.82 at $b$ = 16.2 kpc for a particular
galaxy. 
Subsequently, for the 95 galaxies, the $p_{15.82}$'s are binned according 
to $M_V$, for $M_V$ in the range [-18;-25] with a bin size of 1 mag.. 
For $M_V > -18$, no galaxies were found to display
log(N(SII))$\ge$15.82, and none of the simulated galaxies had $M_V < -25$.
In this way, seven probabilities, $f_{15.82}(M_V)$, are obtained --- these
are displayed in Fig.~\ref{fig:probs} together with the statistical 1-$\sigma$
deviations. As can be seen from the figure, for galaxies of 
$M_V$$\la$$M_V^{\star}$ ($\sim$-23, see below), of order 10\%
of the sight-lines through the galaxies at impact parameter $b$=16.2 kpc
will display a Sulfur II column density of $N(\rm{SII)} \ge 10^{15.82}$ 
cm$^{-2}$ (the value observed for the DLA) --- this is one of the main results 
of this paper. 

The line-of-sight extent or ``thickness'' of the SII distribution can be
roughly estimated as two times the dispersion of the SII number density 
weighted position along the los. Denoting    
\begin{equation}\label{eq:l2}
\bar{l^2} = 
\int_{-l_{\rm{max}}}^{l_{\rm{max}}} l^2 \cdot \it{n}_{\rm{SII}}(l) ~dl ~~, 
\end{equation}
and
\begin{equation}\label{eq:l}
\bar{l} = 
\int_{-l_{\rm{max}}}^{l_{\rm{max}}} l \cdot \it{n}_{\rm{SII}}(l) ~dl ~~, 
\end{equation}
the los position dispersion is calculated as
\begin{equation}\label{eq:sigl}
\sigma_l = \sqrt{\bar{l^2}-\bar{l}^2} ~~. 
\end{equation}
For lines-of-sight of log(N(SII))$\ge$15.82, typical values of the order
$\sigma_l \sim 10$ kpc are found. Hence the thickness of the SII ``layer''
is of the order 20 kpc, but with a considerable variation.

\subsection{Models of the $z$=2.58 galaxy luminosity function}
\label{sec:LFintro}
As shown in the previous subsection, the probability of finding near-solar
metallicity at an impact parameter of $b$ = 16.2 kpc (not surprisingly) 
increases with galaxy luminosity. On the other hand, it is well known that
the number density of galaxies per unit luminosity as a function of luminosity,
$dN(L)/dL$, decreases with luminosity, especially at $L$$\ga$$L_{\star}$,
where the decrease is approximately exponential. To determine the (Bayesian) 
probability distribution of finding log(N(SII))$\ge$15.82 at $b$ = 16.2 kpc
as a function of $M_V$ or $SFR$, one has to fold the two distributions:

Since no luminosity functions (LFs) at $z$=2.58 are available, we shall 
consider three different observational models of the LF, spanning a redshift 
range of $\sim$2.25-3 and a wide range of faint end slopes.  
The first is the $z$$\sim$3 V-band (rest-frame) LF of \cite{sha01} (S01), the 
second is the $z$$\sim$3 V-band (rest-frame) LF of \cite{m07} (M07$z$$\sim$3)
and the third the $z$$\sim$2.25 R-band (rest-frame) LF of \cite{m07}
(M07$z$$\sim$2.25).

As is well known, even at $z \sim 3$, effects of dust extinction on the 
luminosities of galaxies are considerable \citep[e.g.,][]{SLF08}. As the
quantities described in the previous subsection, such as $p_{15.82}(M_V)$ 
depend on true luminosities etc., it is consequently important to correct
the observed luminosity functions for effects of dust extinction. Details
about the luminosity functions and how the extinction corrections are
performed are given in the appendix of this paper, and the reader is
referred to the appendix for all such detail.

\subsection{The constrained probability functions}
\label{subsec:prob}
\begin{figure}
\includegraphics[width=0.48\textwidth]{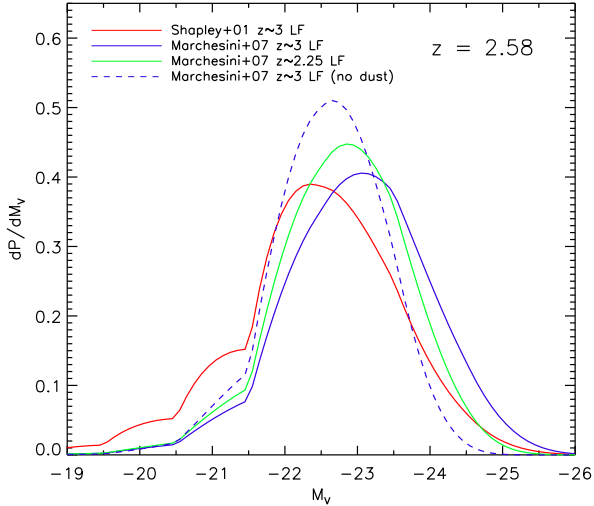}
\caption{Probability distributions, $P(M_V)$, for the three different 
extinction corrected LFs (S01: red curve, M07$z$$\sim$3: blue, solid curve,  
M07$z$$\sim$2.25: green curve). Also shown is the $P(M_V)$ resulting from not
extinction correcting the M07$z$$\sim$3 LF (blue, dashed curve).
\label{fig:PMV}}
\end{figure}
The aim in this section is the following: given that a DLA of log(N(SII))$\ge$
15.82 is located at impact parameter $b$=16.2 kpc relative to a galaxy, what
typical values of $M_V$ and $\log(SFR)$ would the galaxy then be expected to
have, based on the galaxy formation simulations presented in this paper?

\begin{figure}
\includegraphics[width=0.48\textwidth]{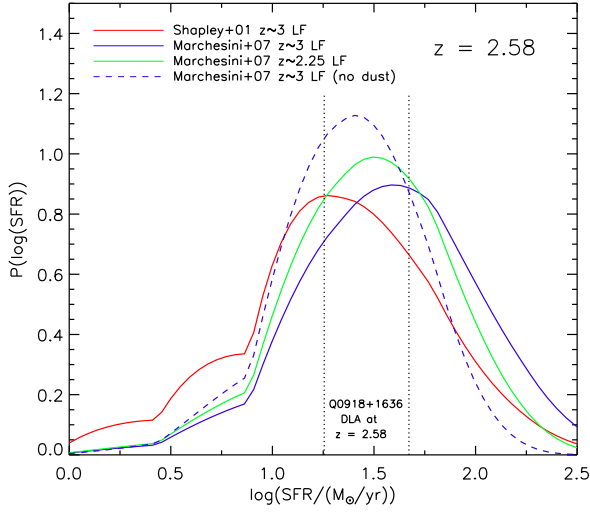}
\caption{Probability distributions, $P(\log(SFR))$, for the three different 
extinction corrected LFs (S01: red curve, M07$z$$\sim$3: blue, solid curve,  
M07$z$$\sim$2.25: green curve). Also shown, is the $P(\log(SFR))$ resulting 
from not extinction correcting the M07$z$$\sim$3 LF (blue, dashed curve.  
Finally shown, by vertical dotted lines, is the observationally estimated $SFR$ 
range for the $z=2.58$ DLA \citep{2013MNRAS.436.361F} 
\label{fig:PSFR}}
\end{figure}

To be quantitative, we calculate the (Bayesian) probability distributions
of $M_V$ and $\log(SFR)$ as follows: For a given small galaxy luminosity 
interval [$L_V$,$L_V+dL_V$], the probability per unit volume, $dP$, of finding
log(N(SII))$\ge$15.82 at impact parameter $b$=16.2 kpc relative to a galaxy
must be proportional to 
\begin{equation}\label{eq:dp}
\frac{dN_{corr}}{dL_V}(L_V)~f_{15.82}(L_V)~dL_V ~~,
\end{equation}
where $f_{15.82}$ was determined in subsection (\ref{subsec:SII}). Changing
variable from $L_V$ to $M_V$, we obtain
\begin{multline}\label{eq:dpmv}
dP \propto \frac{dN_{corr}}{dL_V} \cdot \frac{dL_V}{dM_V} 
~f_{15.82}(M_V)~dM_V = \\
 -\frac{L_V}{2.5 \log(e)} \cdot  
\frac{dN_{corr}}{dL_V}~f_{15.82}(M_V)~dM_V ~~.
\end{multline}
Consequently, under the constraining condition that only $b$=16.2 kpc 
sight-lines of log(N(SII))$\ge$15.82 are considered, the probability 
distribution of DLA galaxy $M_V$ is given by
\begin{equation}\label{eq:pmv}
\frac{dP}{dM_V} =  
\frac{\frac{dN_{corr}}{dL_V}~f_{15.82}(M_V)~L_V(M_V)}
{\int_{-\infty}^{\infty} \frac{dN_{corr}}{dL_V}~f_{15.82}(M'_V)~L_V(M'_V)
~dM'_V} ~~.
\end{equation}
In Fig.~\ref{fig:PMV} the resulting probability distributions are shown for the
three different extinction corrected LFs. To illustrate the effect of the
extinction correction, the probability distribution resulting from not
extinction correcting the M07$z$$\sim$3 LF is also shown. It is seen, that
the probability distributions peak at $L \sim L^{\star}$, which is another
main result of this work. It is also seen that for the S01 LF, the probability
distribution is skewed somewhat towards lower luminosities, due to the
steep faint end slope of the S01 LF. 
\begin{figure}
\includegraphics[width=0.48\textwidth]{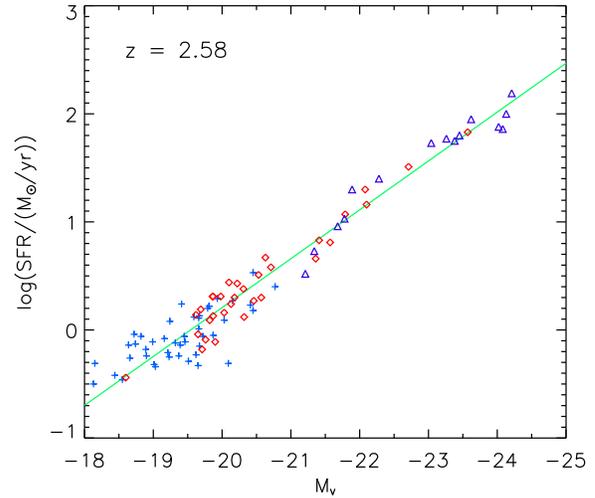}
\caption{Relation between log(SFR) and $M_V$ for the simulated galaxies at
$z$=2.58. The rescaled HR halo simulations are shown by dark blue triangles 
(15 galaxies), the unscaled HR simulations by red squares (32 galaxies) and
the HR disk galaxy simulations by light blue crosses (48 galaxies). Also shown
is a linear regression fit to the data.
\label{fig:SFR_MV}}
\end{figure}
For comparison to observations, it is also convenient to express the 
probability functions as functions of galaxy SFR. In Fig.~\ref{fig:SFR_MV}
the relation between SFR and $M_V$ is shown for the 95 HR sample galaxies.
It is seen, that there is a quite tight relation between the two quantities.
By linear regression one obtains 
\begin{equation}\label{eq:SFRMV}
\log(SFR) = -0.452 M_V - 8.84 ~~,
\end{equation}
where $SFR$ is expressed in units of $\Msun/yr$. This corresponds to
$SFR \propto L_V^{1.13}$, so the two quantities are approximately linearly
related over the range of $M_V$ considered. Applying this transformation,
the probability functions can be expressed as $P(\log(SFR)$) --- the 
transformed probability functions are shown in Fig.~\ref{fig:PSFR}. Also
shown is the observationally estimated range of $SFR$ for the actual
$z=2.58$ DLA \citep{2013MNRAS.436.361F}. As can be seen, the probability
functions all peak in the observational $SFR$ range - this is also an
important results of this work.
To be more quantitative, the probability of finding an $SFR$ in the range
10-100 $\Msun/yr$ is determined. For the extinction corrected LFs, this
probability is 0.65, 0.75 and 0.78. For the non extinction corrected 
M07$z$$\sim$3 LF, the corresponding probability is 0.79. Hence, a central
result of this paper is that, based on the new galaxy formation models,
a) extreme galaxies are not required to explain the observed $z=2.58$ DLA,
and b) good agreement with the observed DLA is found.

\begin{figure}
\includegraphics[width=0.48\textwidth]{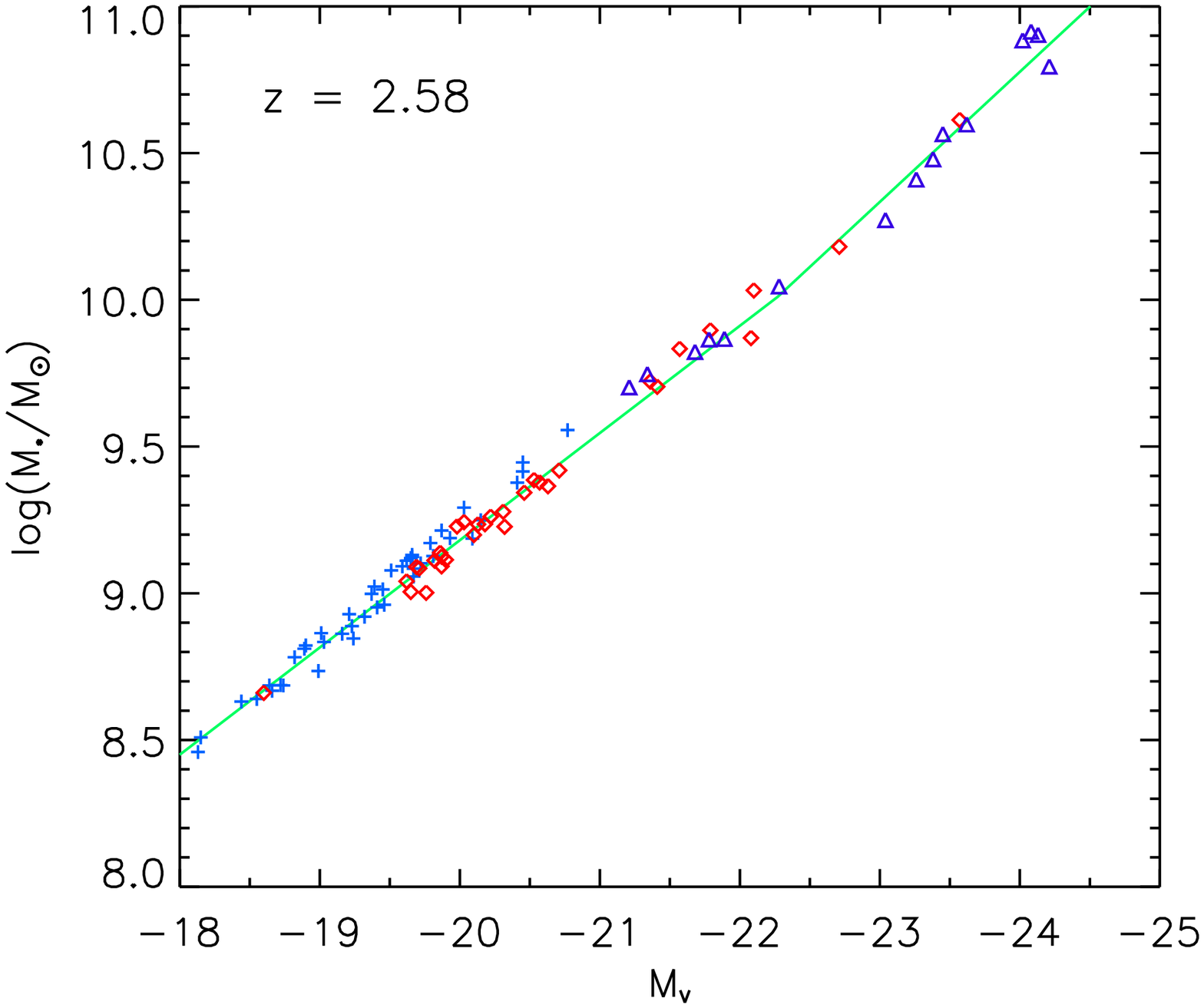}
\caption{Relation between $\log(M_{\star})$ and $M_V$ for the simulated 
galaxies at
$z$=2.58. The rescaled HR halo simulations are shown by dark blue triangles 
(15 galaxies), the unscaled HR simulations by red squares (32 galaxies) and
the HR disk galaxy simulations by light blue crosses (48 galaxies). Also shown
is a two-component linear fit to the data.
\label{fig:mstar_MV}}
\end{figure}
For additional comparison to observations, it is also convenient to express 
the probability functions as functions of galaxy stellar mass. In 
Fig.~\ref{fig:mstar_MV}
the relation between $M_{\star}$ and $M_V$ is shown for the 95 HR sample 
galaxies. It is seen, that there is a quite tight relation between the two 
quantities. The relation can be well fitted by a two component linear relation 
shown in the figure:
 \begin{equation}
\log(M_{\star}) = 
\begin{cases}
-0.3643 M_V + 1.893,   & M_V \ge -22.27, \\
-0.4462 M_V + 0.068,   & M_V < -22.27, \\
\end{cases}
\end{equation}
\begin{figure}
\includegraphics[width=0.48\textwidth]{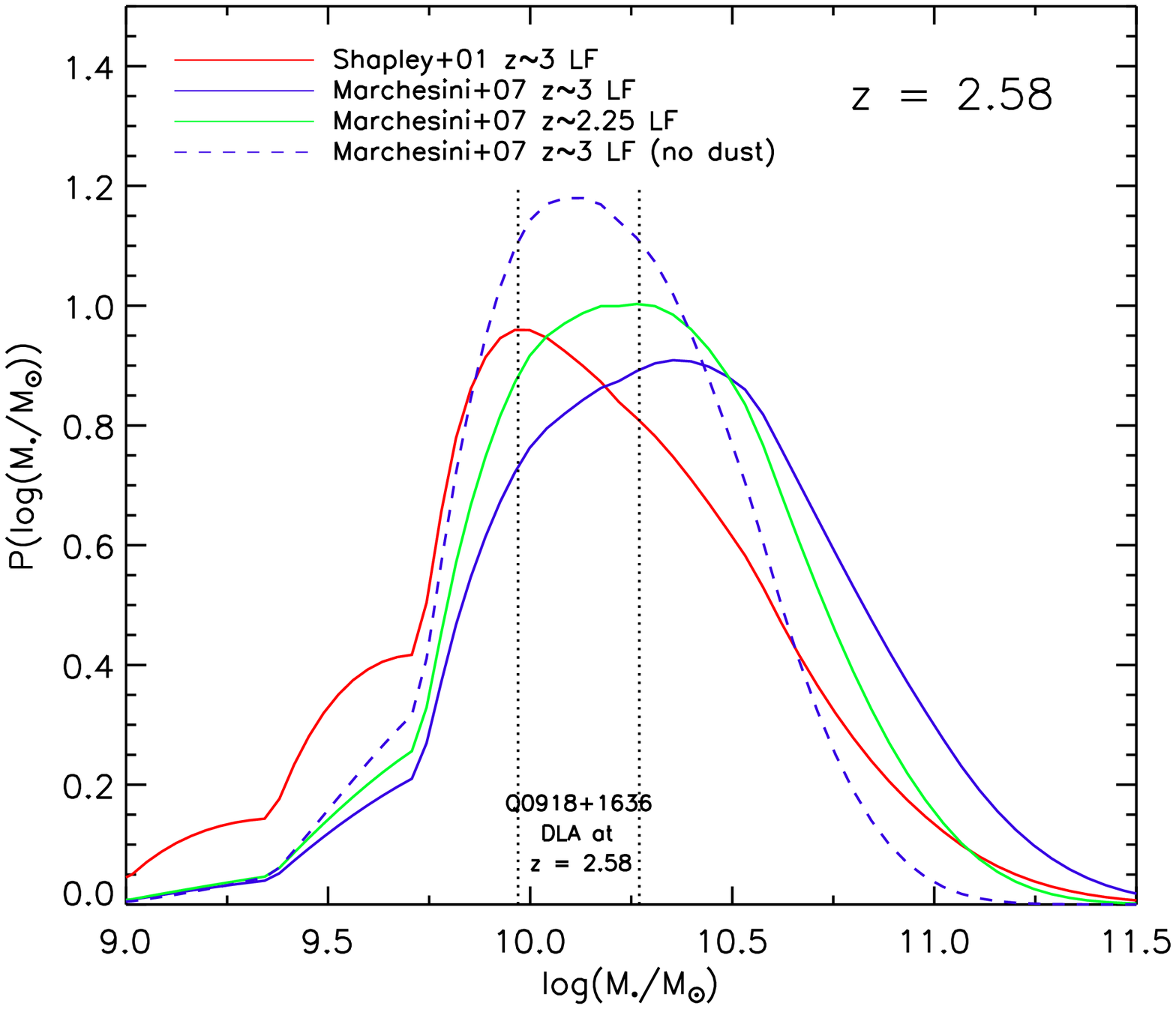}
\caption{Probability distributions, $P(\log(M_{\star}))$, for the three 
different 
extinction corrected LFs (S01: red curve, M07$z$$\sim$3: blue, solid curve,  
M07$z$$\sim$2.25: green curve). Also shown, is the $P(\log(M_{\star}))$ 
resulting 
from not extinction correcting the M07$z$$\sim$3 LF (blue, dashed curve.  
Finally shown, by vertical dotted lines, is the observationally estimated 
1-$\sigma$ range for the $z=2.58$ DLA \citep{2013MNRAS.436.361F} 
\label{fig:Pmstar}}
\end{figure}
Applying this transformation,
the probability functions can be expressed as $P(\log(M_{\star}))$ --- the 
transformed probability functions are shown in Fig.~\ref{fig:Pmstar}. Also
shown is the observationally estimated range of $M_{\star}$ for the actual
$z=2.58$ DLA \citep{2013MNRAS.436.361F}. As can be seen, the probability
functions all peak in or near the observational $M_{\star}$ range.
\begin{figure}
\includegraphics[width=0.48\textwidth]{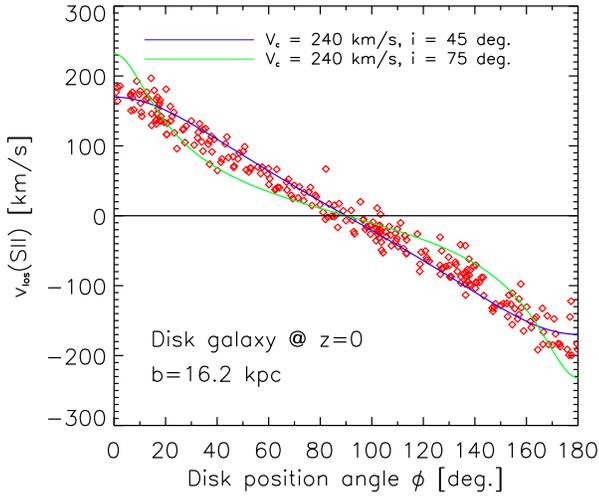}
\caption{For an ideal, plane disk in circular rotation with constant rotation
velocity of 240 km s$^{-1}$ is shown the expected $v_{\rm{los}}(\phi)$ 
(eq.~(\ref{eq:vlos})) for inclination angles $i = 45$ and 75 deg. Also shown,
for a $z=0$ disk galaxy with an extended gas disk, are n(SII) weighted 
line-of-sight velocities for the subset
600 $b$=16.2 kpc sight-lines which have log(N(SII))$\ge$ 15.82 and are hitting
the disk at angles between 15 and 45 deg., corresponding to inclination
angles between 45 and 75 deg.
\label{fig:kin1}}
\end{figure}
\begin{figure}
\includegraphics[width=0.48\textwidth]{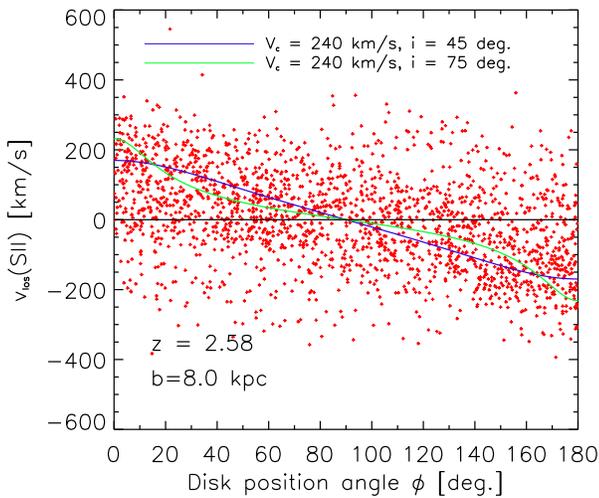}
\caption{For all $z$=2.58 galaxies, which are characterized by having at least
one $b$=16.2 sight-line with
log(N(SII))$\ge$ 15.82, are shown n(SII) weighted line-of-sight velocities for 
the subset of 600 $b$=8.0 kpc sight-lines which have log(N(SII))$\ge$ 15.82 and
are hitting the galaxy at angles between 15 and 45 deg., corresponding to 
inclination angles between 45 and 75 deg. The curves shown are the same
as in Fig.~\ref{fig:kin1}.
\label{fig:kin2}}
\end{figure}

\subsection{Kinematics}
\label{subsec:kin}
From the G{\small ALFIT} analysis \citet{2013MNRAS.436.361F} infer a 
(projected) axis ratio of the DLA galaxy of $b/a=0.43$. The system may be 
described as disk-like, given the elongated shape, and the fact that the 
S\'ersic $n$ close to 1. \citet{2013MNRAS.436.361F} argue that the inclination
angle of the (potential) disk galaxy is about $i\simeq60$ deg.

Moreover, the position of the QSO (and hence the DLA absorbing gas) on the sky 
is located at an angle of about $\phi \sim 45$ deg. relative to
the major axis of the DLA galaxy. The velocity centroid of the 
[\ion{O}{iii}]\,$\lambda$5007 line is
$36\pm20$ km s$^{-1}$ blue-shifted compared to the center of the low-ionization 
absorption lines \citep{2013MNRAS.436.361F}. 

It is clearly of interest to determine whether the simulated galaxies display
signatures of disk-like kinematics, in particular in relation to the
observed DLA galaxy:

\subsubsection{Signatures of disk-like kinematics for the simulated galaxies}
\label{subsubsec:diskkin}
Consider a planar disk with gas in constant circular rotation of velocity
$V_0$, and let the inclination angle with the plane of the sky be $i$ (assumed
in the following to be larger than zero). Denoting the total angular momentum 
of the disk by  $\vec{J}$, a unit vector along $\vec{J}$ will be 
$\vec{j}$=$\vec{J}/|\vec{J}|$. Assuming the x-y plane to be the plane of the 
sky, with origo at the center of the disk, and oriented such that the unit 
vector projected onto the plane of the sky is parallel with the y-axis 
($\vec{j_{x,y}}=(0,j_y$)), then the largest gas line-of-sight recession 
velocity will be observed along the positive x-axis, and the largest 
approaching velocity will be observed along the negative x-axis.

For a given x-y position in the disk's projection onto the plane of the sky, 
we shall denote the absolute value of the angle between the positive x-axis
and the direction from the center to the position by $\phi$ (so $\phi$ is
measured either clock-wise or anti-clock-wise from the positive x-axis, 
depending on the location of the x-y position and takes values between 0 and 
180 degrees). It is then straightforward to show that the line-of-sight
velocity of gas located at ``projected disk position angle'' $\phi$, is given 
by 
\begin{equation}\label{eq:vlos}
v_{\rm{los}} = V_0 \frac{\sin(i)}{\sqrt{1+(\frac{\tan(\phi)}{\cos(i)})^2}} ~~,
\end{equation}
\begin{figure}
\includegraphics[width=0.48\textwidth]{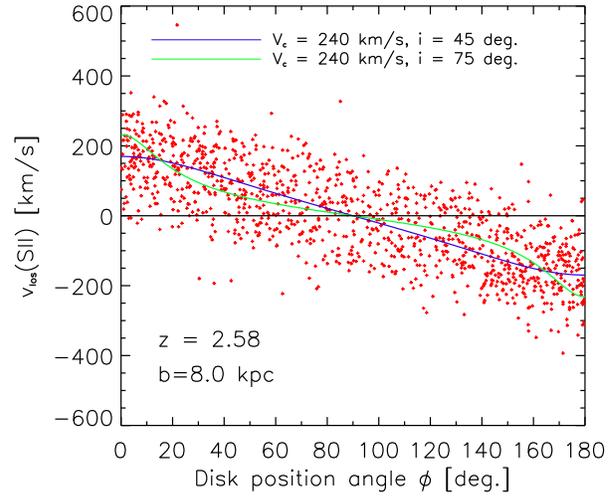}
\caption{Same as Fig.~\ref{fig:kin2}, but restricted to the subset of galaxies 
displaying a young stellar disk {\it and} a gaseous disk-like structure, 
approximately aligned with the stellar disk. 
\label{fig:kin3}}
\end{figure}
\begin{figure}
\includegraphics[width=0.48\textwidth]{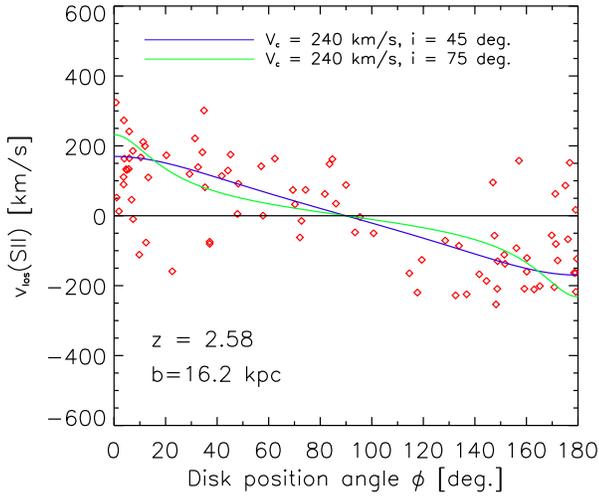}
\caption{Same as Fig.~\ref{fig:kin3}, but now for $b$=16.2 kpc sight-lines.
\label{fig:kin4}}
\end{figure}
First, we test how well the above expression describes what is found
for a $z = 0$ disk galaxy with an extended gaseous disk. The galaxy
has an almost constant rotation curve of amplitude $V_0 \simeq 240$ km 
s$^{-1}$, and is taken from the sample of disk galaxy simulations described in
sec. \ref{subsec:sims}.  As in sec. \ref{subsec:SII}, 600 sight-lines are shot 
through the disk at impact parameter $b = 16.2$ kpc, and the analysis is then
performed in exactly the same way as in that section. In Fig.~\ref{fig:kin1} is
shown a) results of eq.~(\ref{eq:vlos}) for inclination angles $i = 45$ and
75 deg., and b) n(SII) weighted line-of-sight velocities for the subset
of the 600 sight-lines which have log(N(SII))$\ge$ 15.82 and are hitting
the disk at angles between 15 and 45 deg., corresponding to inclination
angles between 45 and 75 deg. As can be seen from the figure, the 
line-of-sight velocities in general fall close to the predictions by
eq.~(\ref{eq:vlos}). Turbulent velocities appear to be at the level
$\sigma \sim 10-20$ km s$^{-1}$, consistent with what is found observationally
for cold $z=0$ disks. 

Next, we analyze the $z = 2.58$ galaxies for similar kinematic signatures.
We first shoot 600 sight-lines at $b = 8$ kpc, to obtain better statistics
and a more clear-cut picture of the kinematic state of the galaxies. The
result for all sight-lines with log(N(SII))$\ge$ 15.82 and for all
galaxies is shown in Fig.~\ref{fig:kin2}. The turbulent velocity dispersion
is now much larger (of the order 100 km s$^{-1}$ --- see below), but there
is still a clear correlation between $v_{\rm{los}}($\ion{S}{ii}) and disk 
galaxy position angle $\phi$ (the two quantities are correlated at the 
20-$\sigma$ confidence level). This hints the presence of rotational motion in 
at least some of the $z=2.58$ galaxies. Third, to elaborate further on this,
all $z = 2.58$ galaxies with log(N(SII))$\ge$ 15.82 sight-lines at $b = 16.2$ 
kpc were visually inspected. The subset of galaxies displaying a young stellar
disk {\it and} a gaseous disk-like structure, approximately aligned with the 
stellar disk were selected --- the stellar disks typically have
an extend of $R \sim 5$ kpc, and the gaseous disks typically extend to
 $R \sim 10$ kpc. In Fig.~\ref{fig:kin3}, we show the relation 
between $v_{\rm{los}}($\ion{S}{ii}) and $\phi$ for this case, which comprises 
56\% of all the sightlines shown in Fig.~\ref{fig:kin2}. The correlation now
stands out very clearly, and is statistically significant at the 41-$\sigma$ 
confidence level.
  
Finally, we perform the analysis of the disk-like galaxies for 
impact parameter $b = 16.2$ kpc. The result of this is shown in 
Fig.~\ref{fig:kin4}. As can be seen, the correlation is still present, but,
as one would expect, not so pronounced at these large galactocentric distances.
The correlation is statistically significant at the 8-$\sigma$ confidence 
level, so we conclude that at least a subset of the galaxies display disk-like
motion even at galactocentric distances $R \ga 23-60$ kpc, depending on the
inclination angle. We note, however, that this result has been obtained in a 
somewhat subjective way, but also that the observed candidate DLA galaxy does
display a disk-like morphology and a S\'ersic $n$ close to 1.

\subsubsection{Kinematical comparison to the DLA galaxy}
\label{subsubsec:DLAkin}
As mentioned above, the gas traced by the low-ion absorption lines is
receding by a line-of-sight velocity of $36\pm20$ km s$^{-1}$ compared
to the velocity centroid of the [\ion{O}{iii}]\,$\lambda$5007 emission
line. The emission originates from the central parts of the DLA galaxy,
and our working hypothesis is that this line traces the systemic velocity
of the DLA galaxy. Moreover, the absorbing gas is located at a disk
position angle $\phi \sim 45$ deg. (or 135 deg. depending on the rotational
direction of the disk-like structure --- see further below).

For comparison to the observed DLA galaxy, we determine the average 
line-of-sight velocity of the SII absorbing gas along the log(N(SII))$\ge$ 
15.82 sight-lines of the ``disk-like'' 
simulated galaxies: For sight-lines located such that $30<\phi<60$ deg., 
we find $\bar{v}_{los} = 102\pm21$ km s$^{-1}$, and a velocity dispersion of
$\sigma=109$ km s$^{-1}$. Considering sight-lines located such that $120<\phi<
150$ deg., one obtains $\bar{v}_{los} = -137\pm23$ km s$^{-1}$, and a velocity 
dispersion of $\sigma=103$ km s$^{-1}$. Combining the two $\phi$-angle 
intervals, exploiting the
symmetry expected in the case of a rotating disk (i.e., changing the sign of 
the $120<\phi<150$ deg. velocities), results in $\bar{v}_{los} = 118\pm15$ km 
s$^{-1}$ and a velocity dispersion of $\sigma=106$ km s$^{-1}$ for $30<\phi<60$
deg. (and $\bar{v}_{los} = -118\pm15$ km and a velocity dispersion of 106 km 
s$^{-1}$ for $120<\phi<150$ deg.). All results refer to inclination angles $i$ 
between 45 and 75 deg., as above.

If the ``disk'' of the DLA galaxy, as a whole, is indeed ``rotating'' away 
from us in the region probed by the DLA absorption with average line-of-sight 
velocity of 118 km s$^{-1}$, then the deviation between this and the 
measured line-of-sight velocity of 36 km s$^{-1}$ amounts to 82 km s$^{-1}$. 
With a velocity dispersion of $\sigma = 106$ km s$^{-1}$, this amounts to 
about 0.8 $\sigma$, so the measured value of 36 km s$^{-1}$ is quite consistent
with the models. If, on the other hand,
the DLA galaxy is ``rotating'' towards us with an average line-of-sight 
velocity of -118 km s$^{-1}$, then the deviation between this and the 
measured +36 km s$^{-1}$ amounts to 154 km s$^{-1}$ or about 1.5 $\sigma$.
Hence, it is about a factor of three more likely that the ``disk'' as a whole
is receding rather than advancing in the region probed by the DLA absorption.
    
For the idealized disk described above, and for $V_0$ = 240 km s$^{-1}$ and
$\phi = 45$ deg., it follows from eq.~(\ref{eq:vlos}) that $v_{\rm{los}} =
98$ km s$^{-1}$ for $i = 45$ deg. and 58 km s$^{-1}$ for $i = 75$ deg. The
average value of $118\pm15$ km s$^{-1}$ and dispersion of 106 km s$^{-1}$
obtained above hence indicates that the circular velocities of the disk-like
galaxies are somewhat larger than 240 km s$^{-1}$ at radii $\sim 15-30$ kpc
from the centers of the galaxies. Indeed, the circular velocities of these 
galaxies are found to lie in the range 290-360 km s$^{-1}$.

\section{Discussion}
\label{sec:Disc}
\subsection{UVB self-shielding and radiative effects of local young stars}
\label{subsec:RT}
Adopting the
mean field approximation of \cite{R13+}, it is assumed in the analysis and in 
the simulations, that the local radiation field, at a given redshift, depends 
mainly on the local 
hydrogen number density $n_H(\vec{x})$. According to \cite{R13+} this is
an excellent approximation, but it is still of value to test other scenarios:
The first alternative considered, is the assumption that effects of neutral
gas shielding can be completely neglected, i.e., assuming that the gas
everywhere is exposed to the full UVB.
This obviously is an extreme case, as medium to high density gas will provide
some shielding of the UVB.
\subsubsection{Neglecting self-shielding of the UVB}
\label{subsubsec:noRT}
To test this case, all HR simulations were run for a period of
100 Myr till $z=2.58$, with UVB self-shielding switched off, i.e. the
gas is assumed to be exposed to the full UVB, independent of gas density.
To make the simulations self-consistent, the radiative cooling and heating 
functions were appropriately modified using Cloudy. So the new simulations
were started from the original simulations at $z=2.67$ and run for the
100 Myr period to $z=2.58$ with the modified simulation code. The 100 Myr
period allows plenty of time for the new ionization balances to be
established. Since all gas in this case is exposed to the full, unimpeded UVB,
higher ionization stages will be more populated for most atoms, including 
Sulfur, compared to the fiducial model. As can be seen from  
\begin{figure}
\includegraphics[width=0.48\textwidth]{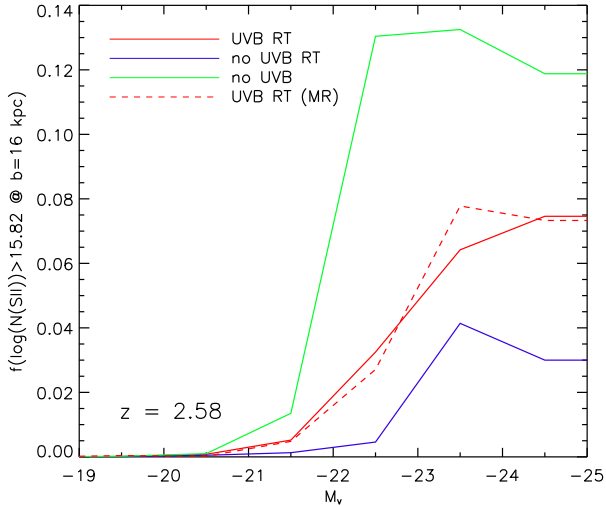}
\caption{For various assumptions about the UVB self-shielding, is
shown the probability of finding log(N(SII))$\ge$15.82 for a sight-line at 
impact parameter $b$ = 16.2 kpc relative to a given galaxy of absolute
magnitude $M_V$ at $z$=2.58. The red solid curve corresponds to the mean field
approximation UVB self-shielding employed in this paper (already shown in Fig.
~\ref{fig:probs} on a log scale), the blue curve corresponds to the case
of no UVB self-shielding, i.e., the gas is everywhere exposed to the
full, unimpeded UVB, and the green curve corresponds to the case of no
H-ionizing UVB at all, i.e., it is assumed that the $E\ge 13.61$ eV part of
the UVB has been completely blocked by neutral hydrogen absorption. 
The red dashed curve corresponds to the fiducial case, but is based on medium
resolution simulations. For more detail, see the text.   
\label{fig:weights_cases}}
\end{figure}
Fig.~\ref{fig:weights_cases}, the fraction of $b$=16.2 kpc 
sight-lines of log(N(SII))$\ge$15.82 per galaxy is indeed lower than for the
fiducial model. In particular, around $M_V \sim -22$ the decrease is very
significant. This implies that the peak of the conditional probability function
is shifted somewhat towards larger luminosities. This is shown in 
Fig.~\ref{fig:PSFR_cases}, where $P(\log(SFR))$ is shown for the   
M07$z$$\sim$2.25 LF in the no self-shielding case, together with the fiducial 
case. 
\begin{figure}
\includegraphics[width=0.48\textwidth]{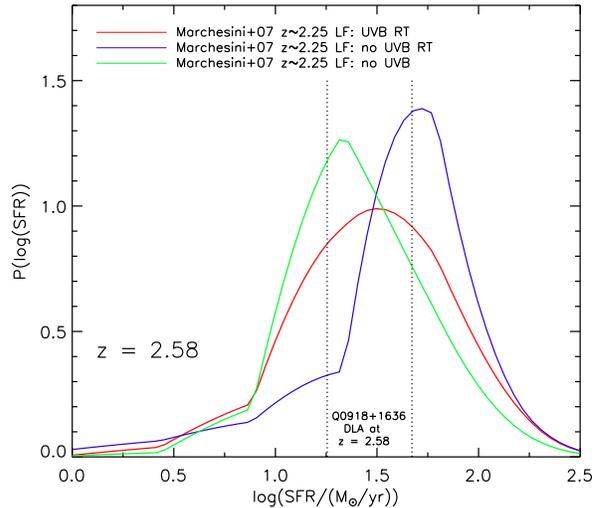}
\caption{Dependence of the probability distributions $P(\log(SFR))$ on the
details of the UVB self-shielding. The extinction corrected  
M07$z$$\sim$2.25 LF is used as an example, and the curves are labeled as in
Fig.~\ref{fig:weights_cases}. 
\label{fig:PSFR_cases}}
\end{figure}

\subsubsection{Neglecting the H-ionizing part of UVB}
\label{subsubsec:noUV}
As another extreme case, we now consider the case where it is assumed that
all UVB H-ionizing photons have been absorbed, such that the gas
is fully shielded from H-ionizing photons. UVB photons of energy less than
13.61 eV are still assumed to be transmitted --- this is important for the
population of SII, since the ionization potential for singly ionizing neutral
Sulfur is 10.36 eV.

To test this case, all HR simulations were run for a period of
100 Myr till $z=2.58$, with H-ionizing UVB switched off, i.e. the
gas is assumed to be fully shielded.
To make the simulations self-consistent, the radiative cooling and heating 
functions were appropriately modified using Cloudy. So, as for the no UVB RT
case,  the new simulations
were started from the original simulations at $z=2.67$ and run for the
100 Myr period to $z=2.58$ with the modified simulation code. Since all gas in 
this case is shielded,
higher ionization stages will be less populated for most atoms, including 
Sulfur, compared to the fiducial model. As can be seen from  
Fig.~\ref{fig:weights_cases}, the fraction of $b$=16.2 kpc 
sight-lines of log(N(SII))$\ge$15.82 per galaxy is higher than for the
fiducial model. In particular, around $M_V \sim -22$ the increase is quite
significant. This implies that the peak of the conditional probability function
is shifted somewhat towards lower luminosities. This is also shown in 
Fig.~\ref{fig:PSFR_cases} (again for the M07$z$$\sim$2.25 LF).
Focusing on the probability of finding an $SFR$ in the range 10-100 
$\Msun/yr$, we find for the M07$z$$\sim$2.25 LF that this probability is 0.82 
for the no UVB self-shielding case and 0.80 for the no
UVB case, so in both cases marginally larger than the 0.78 for the fiducial
case. 

\subsubsection{Radiative effects of young stars in the galaxy}
\label{subsubsec:ys}
Only radiative effects of the UVB are considered in the fiducial model. Even
though an impact parameter of 16.2 kpc at $z=2.58$ implies that the galaxy
is being probed in its outer parts, radiation from the young stars in the
galaxy may in principle still affect the ionization balances in the gas in the
outer parts of the galaxy, in particular radiation of $E<$13.61 eV --- see
below.

Focusing first on H-ionizing radiation, we calculate for each line-of-sight
the minimum distance to any star particle of age less than 10 Myr in the 
galaxy. For galaxies where
$b$=16.2 kpc sight-lines of log(N(SII))$\ge$15.82 are present, the average
over these minimum distances are then calculated for each galaxy. Finally,
the average over all such galaxies is calculated. The result obtained is
$<d_{min,10 Myr}> = 3.7\pm0.3$ kpc. The galaxies for which high metallicity
$b$=16.2 kpc sight-lines are found typically to have disk-like morphology,
with the bulk of the SII absorption arising in or near the ``disk''.
For densities $n_H\sim1$ cm$^{-3}$, typical HII region radii are an
order of magnitude smaller than $<d_{min,10 Myr}>$ (e.g., \cite{MK87}), so it 
is unlikely
that H-ionizing radiation from young stars in the galaxy affects the
ionization balance along the sight-lines probed. In addition, in section 
\ref{sec:outgal} it will be argued that the $\alpha$-elements, like Sulfur,
found in the outer galaxy mainly were produced by evolving stars in the 
inner galaxy, and subsequent transported outward by galactic winds. Hence,
young, UV emitting stars are expected to be located in the inner galaxy
($r\la 10$ kpc --- see section ~\ref{sec:outgal} ), whereas 
the metal-rich gas probed by the $b$=16.2 kpc sight-lines will be located
at galactocentric distances of $r \ga15-30$ kpc.

For radiation of $E<$13.61 eV the situation is quite different, however. Such
radiation is transmitted almost unimpeded through the interstellar gas (apart
from at or near the Lyman line wavelengths). Since the ionization potential
of neutral Sulfur is 10.36 eV, we shall, for the purposes of the present
work, focus on the radiation energy interval 10.36-13.61 eV. For each 
sight-line, we calculate at each ``grid-point'' the total flux at 
$\lambda = 1105$ {\AA} from all stars in the galaxy of age less than 34 Myr
(the lifetime of a 9 $\Msun$ star), using the Starburst99 models \citep{L.99}.
We then calculate, for each sight-line,
the $n_{SII}$ weighted $\lambda = 1105$ {\AA} flux. For all $b$=16.2 kpc 
sight-lines of log(N(SII))$\ge$15.82, we subsequently calculate the average
$\lambda = 1105$ {\AA} flux for the galaxy, and finally we average over
all such galaxies. It is found, that the flux from the young stars is about
twice the flux of the UVB at 1105 {\AA}, so in principle this contribution
could be important for the ionization balance. To investigate this further,
we produced artificial UVBs, where the flux in the energy interval 10.36-13.61
eV was increased by a factor of 100 relative to the Haardt \& Madau (2012)
UVBs. We then proceed as before, using Cloudy to determine SII densities etc.
It is found, that the results of this are almost indistinguishable from the
fiducial case. We conclude that the effects of radiation from young stars
in the galaxies are unimportant for the purposes of this work.

\subsection{Non-local SII absorption}
\label{subsec:nonloc}
The galaxies considered in this work have all been simulated using the
``zoom-in'' technique (sec.~\ref{subsec:sims}). The advantage of the technique
is that high resolution is achieved in and near the galaxy --- the 
disadvantage, however, is that the galaxy region is only well resolved out to
$\sim$100-200 kpc from the center of the galaxy. This can potentially be 
problematic, since a non-local galaxy, of lower luminosity and with the QSO
located at smaller impact parameter than the $b$=16.2 kpc ``candidate 
galaxy'', could in principle be responsible for the DLA absorption. Assuming
that the redshift of the ``candidate galaxy'' is known, e.g., from  
 [\ion{O}{iii}]\,$\lambda$5007 or H$\alpha$ emission, obviously secondary,
non-local galaxies situated sufficiently far along the line-of-sight can
be identified as not associated with the ``candidate galaxy'' from the
velocity difference. But for velocity differences of $\sim$500 km/s or less,
the case is less clear cut. Assuming pure Hubble flow at $z=2.58$, this
would correspond to distances of $\sim$2 physical Mpc, or $\sim$7 co-moving
Mpc. To address the effects of non-local galaxies quantitatively, a 
small-scale cosmological hydro/gravity simulation, previously run to
$z=$2.51 and based on a Salpeter (1955) IMF and a previous SN feedback scheme
(e.g., \cite{SL06}) is utilized. The simulation volume is part of the dark 
matter only cosmological volume described in sec.~\ref{subsec:sims}. The
(cubical)
small-scale cosmological simulation has a box-length of $l$=3.2 physical Mpc or
11.5 co-moving Mpc. As this box-length is somewhat small for the present
purpose, the box is replicated to create a larger cubical box, of box-length
$L$=9.6 physical Mpc or 34.4 co-moving Mpc and consisting of 3$^3$ = 27 of the
original boxes. 

\begin{figure}
\includegraphics[width=0.48\textwidth]{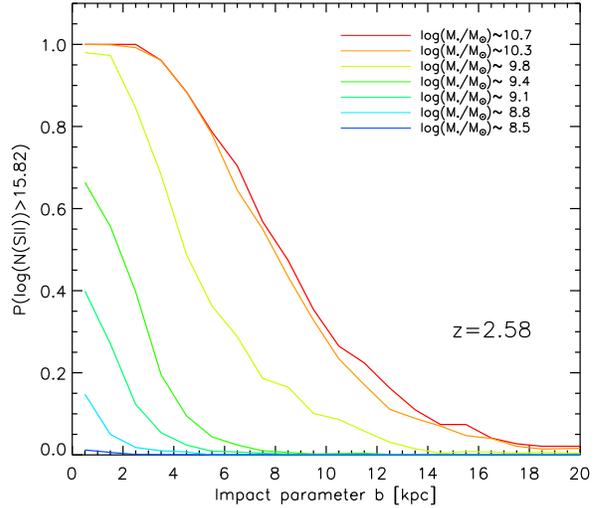}
\caption{Probability function $P_{15.82}(b,M_{\star})$ shown for 7 different
values of stellar galaxy mass, $M_{\star}$.
\label{fig:P1582}}
\end{figure}

To determine the probability of detecting log(N(SII))$\ge$15.82 as a function
of impact parameter, for each of the presently simulated galaxies, 200 
sight-lines are shot randomly through each galaxy at impact parameters
$b = 1, 2, 3, ..., 25$ kpc. The galaxies are then binned according to stellar
mass, and a probability function $P_{15.82}(b,M_{\star})$ is constructed. The
probability function thus obtained is shown in Fig.~\ref{fig:P1582} for 7 
different stellar galaxy masses.
For each of the 589 galaxies located in the central $l$=3.2 physical Mpc box,
1000 lines-of-sight are shot through the full, replicated box, each at an 
impact parameter of $b$=16.2 kpc relative to the galaxy. The directions of the 
1000 lines-of-sight are chosen at random, and lines-of-sight which intercept
a brighter non-local (see below) galaxy at impact parameter $b\le$16.2 kpc are 
rejected (see \ref{subsec:SII}). In this way, for each of the
589 galaxies and for each line-of-sight, the quantity 
\begin{equation}\label{eq:tau1582}
\tau_{15.82} =
\sum\limits_{i=1}^n P_{15.82}(b_i,M_{\star,i})~~, 
\end{equation}
is determined, where the sum is taken over all galaxies in the large box, 
situated at distances of 200 kpc or more from the galaxy under consideration 
and intercepted at $b \le 25$ kpc. In practice, it turns out that for any
of the sight-lines probed, at most one such non-local galaxy in the replicated 
box contributes with $N(SII)\ge 10^{15.82}$ cm$^{-2}$. 

Subsequently, for each of the 589 galaxies the average of $\tau_{15.82}$ over 
the 1000 lines-of-sight, $<$$\tau_{15.82}$$>$, is calculated, and finally,
by binning the 589 galaxies according to $M_V$, the probabilities of
non-local SII absorption, $p_{\rm{15.82,non-local}}(M_V)$ are determined. To
obtain a larger dynamical range, the calculations are repeated with the 
replicated box scaled by a (linear) factor of 1.6, and galaxy masses scaled
by a factor of 1.6$^3$. The resulting non-local SII absorption probability
functions are displayed in Fig.~\ref{fig:probs}. As can be seen from the 
figure, for the relevant values of $M_V$, the probability of non-local
SII absorption is 2-3 orders of magnitude less than the probability of 
absorption in or near the galaxy itself. To first order, it is hence justified
to exclude effects of non-local absorption from the analysis. The average
path-lengths for the unscaled and scaled replicated boxes are 40 and 66 cMpc,
respectively. Assuming pure Hubble flow, this corresponds to velocity shifts
of about 3000 and 4800 km/s, respectively, which should be easily detectable.
Hence, the box-sizes used for the non-local effects analysis seem sufficiently
large.

\subsection{Effects of possible Sulfur depletion into dust grains}
\label{subsec:depletion}
In this subsection, effects of (hypothetical) depletion of Sulfur into dust
grains is discussed:
Sulfur is only weakly depleted into dust grains (e.g., \cite{FS97}, but 
recent work indicates that some amount of depletion may take place --- see, 
e.g., \cite{C.09} and \cite{J09}. 

\citet{2011MNRAS.413.2481F} obtain log(N(SII)) = 15.82$\pm$0.01, and 
log(N(ZnII)) = 13.40$\pm$0.01 for the DLA. This corresponds to 
``metallicities'' of
[SII/H]=-0.26$\pm$0.05 and [ZnII/H]=-0.12$\pm$0.05, respectively. As Zinc
can be assumed to be un-depleted by dust, this could be taken to indicate a 
small amount of Sulfur depletion, although, given the uncertainties, the
effect is small. As Sulfur is an $\alpha$-element, we moreover 
(conservatively) assume that [$\alpha$/Fe]=0.24 at $z$=2.58 (based on the
simulations). Assuming that Zn traces iron, this all in all implies that 
log(N(SII)) = 16.20 should be a conservative upper limit to the SII
column density of the DLA. 

\begin{figure}
\includegraphics[width=0.48\textwidth]{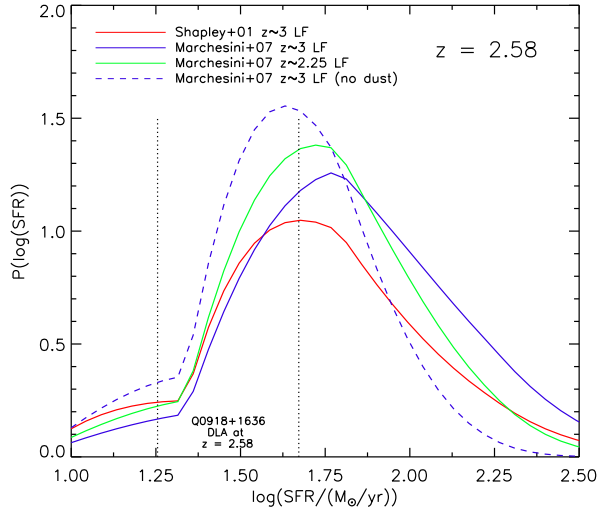}
\caption{Constrained probability distributions of galaxy SFR for the case 
log(N(SII))$\ge$16.20 and $b$=16.2 kpc --- symbols as in Fig. \ref{fig:PSFR} 
\label{fig:PSFR16.20}}
\end{figure}

The analysis described in sections \ref{subsec:SII} and \ref{subsec:prob} is
then re-performed, but now for log(N(SII))$\ge$16.20, rather than 15.82. The
resulting constrained probability distributions of galaxy SFR are shown in
Fig. \ref{fig:PSFR16.20}. As can be seen from the figure, the peaks of the
probability functions move towards larger SFR, but they are still located
close to the observed SFR of the proposed DLA galaxy counterpart. Moreover,
the probability of finding an $SFR$ in the range
10-100 $\Msun/yr$ is 0.85, 0.79 and 0.88 for the three extinction corrected
LFs considered. For the non extinction corrected M07$z$$\sim$3 LF, 
the corresponding probability is 0.94. Hence, the main results
obtained in section \ref{subsec:prob} still hold, if the ``true'' SII
column density is 10$^{16.20}$ cm$^{-2}$.  

\subsection{Statistical effects}
\label{subsec:stat}
Given that the probabilities $f_{15.82}(M_V)$ shown in Fig.~\ref{fig:probs}
are based on 95 sample galaxies, the constrained probability functions
derived in \ref{subsec:prob} will be affected by statistical effects. In order
to assess the importance of these effects we carried out 1000 Monte-Carlo
simulations. Each of the seven $f_{15.82}(M_V)$ were assumed to be normally
distributed with mean values and dispersions set to the values inferred from
the 95 sample galaxies, as shown in Fig.~\ref{fig:probs}. For each MC 
realization of the $f_{15.82}(M_V)$'s, the constrained probability functions
were constructed, and on the basis of these, the average SFRs corresponding
to this realization were determined for each of the luminosity function
models. Based on the 1000 MC realizations, the average SFRs determined were 
36, 56, 43 and 32 $\Msun/yr$, with
dispersions of 4, 6, 4 and 3 $\Msun/yr$, for the S01, M07$z$$\sim$3, 
M07$z$$\sim$2.25 and
M07$z$$\sim$3 (no dust-correction) LFs, respectively. For the actual set of
$f_{15.82}(M_V)$'s (Fig.~\ref{fig:probs}) and corresponding
probability functions (Fig.~\ref{fig:PSFR}), we find average SFRs
of 36, 57, 42 and 32 $\Msun/yr$ for the four LFs, respectively. Hence, there
is no indication that statistical effects affect the main conclusions obtained
in this paper. 

\subsection{Numerical resolution}
\label{subsec:numres}
It is important to check that the results obtained do not depend on the
resolution of the numerical simulations. To address this, we go through
the analysis described in section \ref{sec:results}, using the MR simulations
rather than the HR simulations. The MR simulations have four times lower
mass resolution than the HR simulations. As can be seen from 
Fig.~\ref{fig:weights_cases}, the fraction of $b$=16.2 kpc sight-lines of 
log(N(SII))$\ge$15.82 per galaxy as a function of $M_V$ is quite similar to
what is found for the fiducial model. To quantify this, probability functions 
are determined on the basis of the MR simulations, as described in section
\ref{sec:results}. These are used to calculate average SFRs for the
four cases considered in the previous subsection. One obtains
38, 59, 45 and 34 $\Msun/yr$ , quite close to what is found for the
HR simulations.

\subsection{Lack of AGN feedback}
\label{subsec:agn}
The present simulations do not include effects of feedback from supermassive 
black holes (BHs) --- below we comment briefly on this possible shortcoming.  

The formation and evolution of BHs is thought to be 
linked to galaxy evolution.  Correlations between black hole mass and galaxy 
properties indicate a considerable degree of co-evolution 
\citep[e.g][]{richstone98,ferrarese00,gebhardt00,marconi03,haring04}.  
The energy released by Active Galactic Nuclei (AGNs) is likely liberated 
in connection with accretion of matter onto the BH. AGN hence provide a
physical link between black hole growth and the broader galaxy
\citep[e.g.][]{silk98,dimatteo05}.  Yet the physical processes of black hole
fueling that trigger AGN are not fully understood.

Simulations show that major mergers funnel gas toward the central black holes
\citep[][]{hernquist89, barnes92, dimatteo05}, but although mergers probably do
enhance BH activity \citep{ellison11}, observations have not strongly 
supported the major merger scenario for most AGNs: AGNs are typically  located 
in isolated galaxies, with no obvious signs of a major merger 
\citep{grogin05, pierce07, gabor09, georgakakis09}.  

As AGNs are common in star-forming disks \citep{hwang10, cisternas11, 
kocevski12, schawinski12,juneau13}, it is of interest to study the interplay
between the star-forming gas in (gas-rich) disk galaxies and a central BH
using very high numerical resolution in order to resolve as well as possible
the processes occurring in the central galaxy. \cite{GB13} performed such a
study using AMR (adaptive mesh refinement) at a resolution of 6 pc. They
focused on the thermal feedback from the AGN, and although the AGN drives
uni/bi-polar outflows vertical to the disk at the center of the galaxy 
(expansion along the path of 
least resistance, cf. \cite{FQ12}) the star-formation in the disk in general
is only very weakly affected. \cite{R.15} followed up on this work, and showed
that taking in addition also the radiation feedback from the AGN into account
(even at quasar luminosities) still only affects the general star-formation
in the disk very little. To be conservative, the authors even neglected
the partial absorption of UV and X-ray photons by the (putative) dusty torus. 
\cite{NK13} reached similar conclusions to the above based on their
simulations of AGN feedback in isolated galaxies.

Taken at face value, these results could hint that the omission of AGN feedback
in the current work is not a major limitation. However, as pointed out by
the authors, they study isolated galaxies, and AGN feedback might still affect
galaxy evolution in a cosmological context. Also, the apparent ``quenching''
of late time star formation observed for some (typically massive) galaxies
(``red and dead'' galaxies) may be interpreted most simply as an effect of AGN 
feedback \citep[e.g.][]{B.06,C.06}.

We plan to include effects of AGN feedback in future work.

\section{How was the high metallicity gas in the outer galaxy enriched?}
\label{sec:outgal}
As has been argued in this work, the SII absorbing gas is most likely 
associated with the DLA galaxy. It is hence of considerable interest to
investigate how this metal rich gas, located at galactocentic distances of 
$\sim 15-30$ kpc, was enriched. Possible mechanisms include a) the gas 
is located in a disk like structure, already in place at $z$=2.58, and has
been enriched locally through {\it in situ} star formation, b) the metal rich
gas has been stripped off passing satellite galaxies, and c) the enriched
gas has been deposited in the outskirts of the galaxy through the effects of 
metal enriched winds ejecting gas and metals from the inner parts of the DLA
galaxy. In the following it will be argued that mechanism ``c'' is by far 
the most important.

Due to the Lagrangian nature of SPH, it is possible to trace
a given fluid element (SPH particle) back in time. Hence, the
position history, thermal history, enrichment history etc. can be determined
for the fluid element. The time resolution will be set by how often the
simulation state is saved --- in the present case, once per 100 Myr. In the
following, this capability of determining the history of each SPH particle will 
be used to determine the position of SPH particles of interest during their
major metal enrichment episodes. 

To facilitate the analysis, it will be restricted to SPH particles that are rich 
in SII at $z$=2.58. Specifically, only SPH particles with an abundance of {\it singly
ionized} Sulfur exceeding the (total) solar Sulfur abundance, i.e.
\begin{equation}\label{eq:highabun}
Z_{\rm{S}} \cdot f_{\rm{SII}} > Z_{\rm{S},\odot} ~,
\end{equation}
are considered in the analysis. It may seem overly restrictive to consider only
such very SII-rich particles, but it turns out that for log(N(SII))$\ge$ 15.82
sight-lines at $b = 16.2$, $\sim$ 75\% of the contribution to N(SII) 
(eq.~(\ref{eq:NSIIv4})) stems from such particles. Moreover,
for illustrative purposes, the analysis is carried out for just one $z$=2.58
galaxy, of $p_{15.82} \sim 0.05$ and $M_V = -24.02$, but the results for other 
galaxies are similar. 
\begin{figure}
\includegraphics[width=0.48\textwidth]{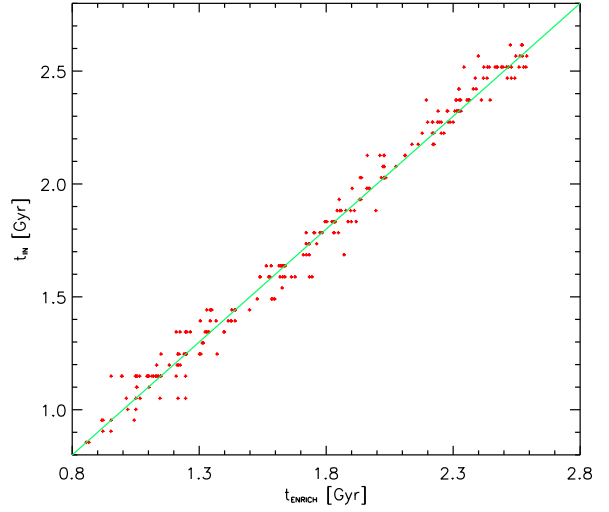}
\caption{Relation between ``inner galaxy time'', $t_{\rm{in}}$, and ``Sulfur
enrichment time'', $t_{\rm{enrich}}$, shown by red symbols. Also shown, by 
the green line,
is the relation $t_{\rm{in}} = t_{\rm{enrich}}$. For more detail, see the text.
\label{fig:tt}}
\end{figure}

For the galaxy at $z$=2.58, the Sulfur enrichment histories for all SPH 
particles located at galactocentric distances between 15 and 30 kpc, and 
satisfying the above SII-abundance constraint, are determined. Moreover, 
the positions of the particles relative to the proto-galaxy are determined in 
all relevant time-frames prior to the $z$=2.58 frame (it turns out that, over 
the time span of interest, the last major merger takes place at $t \sim 0.7$ 
Gyr, so the main proto-galaxy is well defined at later times).

For each such SPH particle, two characteristic times are determined: 
1) The Sulfur enrichment time, $t_{\rm{enrich}}$, measures the average
time of enrichment of the particle, and is defined as 
\begin{equation}\label{eq:tenrich}
t_{\rm{enrich}} =
\sum\limits_{i}^{} \Delta Z_S(t_i)~t_i ~~, 
\end{equation}
where $\Delta Z_S(t_i)=Z_S(t_{i+1})-Z_S(t_i)$, $t_i$ is the time
since Big Bang of frame $i$, and the sum in most cases runs over 20 
time-frames dumped at $t$ = 0.7 Gyr, 0.8 Gyr, ..., 2.6 Gyr, but see below. 
2) The inner galaxy time, $t_{\rm{in}}$, measures the average time at 
which the particle is located in the inner galaxy, and is defined as 
\begin{equation}\label{eq:tin}
t_{\rm{in}} =
\frac{1}{N}\sum\limits_{j=j_{\rm{min}}}^{j_{\rm{max}}} t_j ~~, 
\end{equation}
where the sum runs over the number of consecutive frames, 
$N=j_{\rm{max}}-j_{\rm{min}}+1$, for which 
the galactocentric distance of the SPH particle satisfies $r < r_{\rm{in}}$.
The "inner galaxy radius" is set to $r_{\rm{in}} = 10$ kpc, but the main 
results are not sensitive to moderate variation of this parameter. It turns
out that 97\% of the SII-rich SPH particles, selected as described above,
have been in the inner galaxy from one and up to four times prior to the
$z$=2.58 frame. Due to the limited time-resolution of 100 Myr, it is even
possible that some fraction of the remaining 3\% of SPH particles have also
been in the inner galaxy at some point.   
 
Fig.~\ref{fig:tt} shows the relationship between $t_{\rm{in}}$ and 
$t_{\rm{enrich}}$. As can be seen from the figure, the two quantities are
strongly correlated. Some of the SPH particles have been located in the inner 
galaxy two or even three times --- for these particles, the summation in 
eq.~(\ref{eq:tenrich}) is performed over the appropriate range of time frames,
and two or three data points are shown in Fig.~\ref{fig:tt} for such particles.

\begin{figure}
\includegraphics[width=0.48\textwidth]{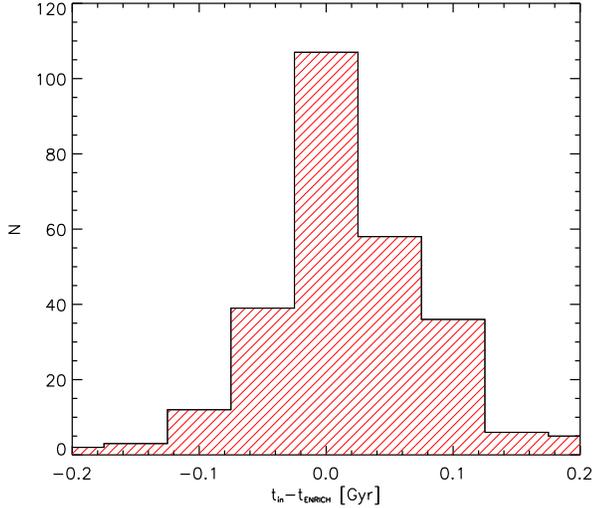}
\caption{The distribution of $t_{\rm{in}} - t_{\rm{enrich}}$. 
\label{fig:tth}}
\end{figure}

To illustrate the correlation further, in Fig.~\ref{fig:tth} is shown the
distribution of $t_{\rm{in}}-t_{\rm{enrich}}$ as a histogram. As can be 
seen from the figure, the width of the distribution is comparable to the time
resolution (100 Myr), indicating that the correlation between $t_{\rm{in}}$ and
$t_{\rm{enrich}}$ is indeed very tight.

The above correlation strongly suggests that the bulk of metal rich SPH 
particles 
located in the outer galaxy ($15<r<30$ kpc) at $z$=2.58 have been enriched 
in the inner galaxy. Subsequently, the particles have been deposited in the 
outer galaxy, likely due to the effects of galactic winds: $\alpha$-elements 
like Sulfur are predominantly produced in core collapse (type II) supernovae, 
and as the SNII explode, metals as well as large amounts of thermal and
kinetic energy are injected into the ISM, potentially driving 
galactic outflows.  

\begin{figure}
\includegraphics[width=0.48\textwidth]{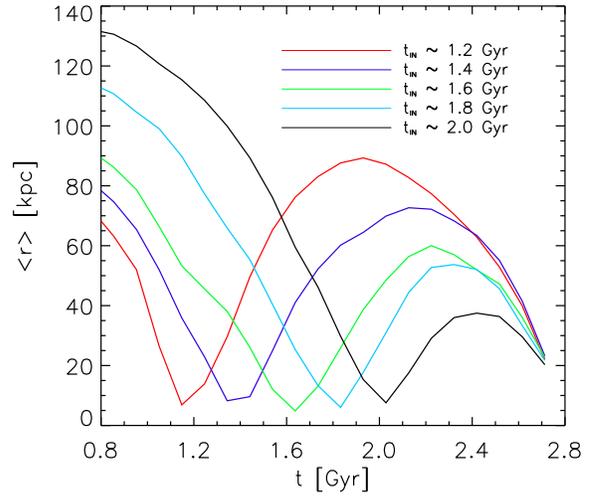}
\caption{Average galactocentric distance, $<r>$, of the
selected metal rich SPH particles binned according to the value of 
$t_{\rm{in}}$ --- for clarity the analysis has been restricted to SPH
particles that are only once in the inner galaxy during the time interval
under consideration.
\label{fig:rt}}
\end{figure}

To further substantiate the galactic wind scenario, shown in 
Fig.~\ref{fig:rt} is the average galactocentric distance, $<r>$, of the
selected metal rich SPH particles for $t_{\rm{in}}$ in five bins:
$t_{\rm{in}}$ = 1.1-1.3 Gyr, 1.3-1.5 Gyr etc.     
As can be seen from the figure, the gas is initially accreted onto the
inner proto-galaxy, then expelled to larger radii and eventually 
re-accreted to $15<r<30$ kpc at $z$=2.58. This part of the analysis
has been restricted to the subset of particles, which have only been once
in the inner galaxy (comprising 71 \% of the SPH particles under 
consideration, i.e., the bulk of the SPH particles), but a similar
behavior is found for the SPH particles that are in the inner galaxy on 
multiple occasions.   

\begin{figure}
\includegraphics[width=0.48\textwidth]{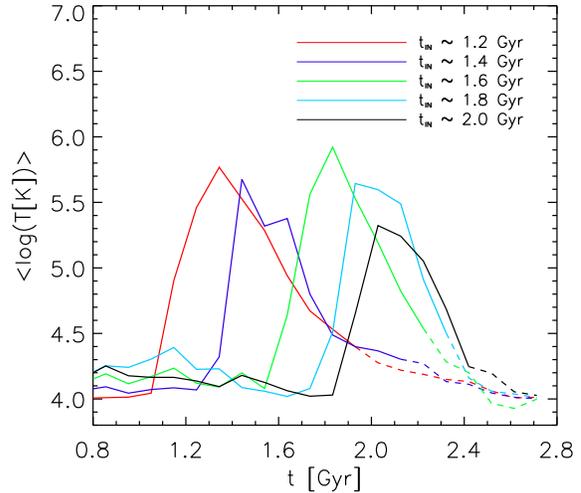}
\caption{Median temperature of SPH particles in the same $t_{\rm{in}}$ 
bins as in Fig.~\ref{fig:rt}. Dashed lines indicate that the SPH particles
are in the second accretion phase --- see text for details. 
\label{fig:tempt}}
\end{figure}

In Fig.~\ref{fig:tempt} is
shown the median temperature of the SPH particles in the above bins. As
can be seen, the gas is initially accreted at a fairly low temperature,
$\log(T) \sim 4.2$, and then, at $t \simeq t_{\rm{in}}$, heated to 
$\log(T) \la 6$. Subsequently, as the gas expands (Fig.~\ref{fig:rt}),
it cools and the density drops to $n_H \sim 10^{-4}-10^{-3}$ cm$^{-3}$.
Eventually, the gas starts re-contracting (this phase is marked by dashed
lines in Fig.~\ref{fig:tempt}). Despite effects of compressional heating, the 
temperature continues decreasing (from $\log(T) \simeq 4.5$ to 
$\log(T) \simeq 4.0$). 
This is due to the radiative cooling rate being larger than the compressional
heating rate, and the temperature balance being set rather by the UVB heating
rate. As the density keeps increasing during the re-accretion process, the
effect of UVB heating keeps decreasing, and the gas temperature consequently 
drops.

The focus here has been on Sulfur, which is an $\alpha$-element, primarily
produced by SNIIs. Iron peak elements are also produced abundantly by SNIa's, 
on a considerably larger stellar evolutionary timescale of the order 1 Gyr  
\citep[e.g.,][]{L.02}. However, the timescale is sufficiently short compared
to the age of the Universe at $z$=2.58, that the outer galaxy may have
been enriched by iron from stars exploding as SNIa {\it in situ} in the
outer galaxy. Performing the same analysis as described above (for Sulfur)
and restricting the analysis to SPH particles of [Fe/H]$>$-0.25 (this
results in approximately the same number of SPH particles for the analysis)
it is found that 90\% of the particles have been located in the inner galaxy
one or more times. This indicates that the bulk of the metal rich gas in
the outer galaxy was enriched in the inner galaxy and transported outward
by winds, but that some additional SNIa enrichment may have taken place
{\it in situ}. Also, if $t_{\rm{enrich}}$ is based on iron rather than
Sulfur, then the relation between $t_{\rm{in}}$  and $t_{\rm{enrich}}$
is not as tight as the one displayed in Fig.~\ref{fig:tt}. This again 
indicates that some {\it in situ} enrichment by SNIa's may have occurred. 

\section{Conclusion}
\label{sec:concl}
It is found that a) for $L$$\sim$$L_{\star}$ galaxies (at $z$=2.58), about 10\%
of the sight-lines through the galaxies at impact parameter $b$=16.2 kpc
will display a Sulfur II column density of $N(\rm{SII)} \ge 10^{15.82}$ 
cm$^{-2}$ (the observed value for the DLA), and 
b) considering only cases where a near-solar metallicity will be detected at
16.2 kpc impact parameter, the (Bayesian) probability distribution of galaxy
star formation rate (SFR) peaks near the value actually observed for the
DLA galaxy counterpart of $27^{+20}_{-9} \Msun$/yr. The probability
that the SFR lies in the range 10-100 $\Msun$/yr is found to be 0.65-0.80
for a very broad range of assumptions about the $z$=2.58 galaxy luminosity
function.

Hence, the main result is that extreme galaxies are not required to match
the high impact parameter and high metallicity DLA. In fact, very good 
agreement between the theoretical predictions and the observed properties of 
the proposed DLA galaxy counterpart is found. It is shown that this result 
is very robust, as it is insensitive to large variations in the treatment of 
UVB self-shielding, statistical uncertainties and possible effects of dust 
depletion.
It is also shown that effects of non-local absorption are unimportant, and
the result is shown to hold for a wide range assumptions about the $z$=2.58 
V-band luminosity function. Finally, the result is found to be robust to
a change of a factor of 4 in the (numerical) mass resolution of the 
simulations. 

It is argued, that the bulk of the $\alpha$-elements, like Sulfur, 
traced by the high metal column density, $b$=16.2 kpc absorption lines, have 
been produced by evolving stars in the inner galaxy, and subsequent 
transported outward by galactic winds.

\section*{Acknowledgements}
We are grateful to the referee, Simeon Bird, for a number of suggetstions and
comments, that improved the presentation of this work.

The research leading to these results has received funding from the European
Research Council under the European Union's Seventh Framework Program
(FP7/2007-2013)/ERC Grant agreement no.  EGGS-278202.

The simulations were performed on the facilities provided by HPC/UCPC at
University of Copenhagen.

\def\aj{AJ}
\def\araa{ARA\&A}
\def\apj{ApJ}
\def\apjl{ApJL}
\def\apjs{ApJS}
\def\apss{Ap\&SS}
\def\aap{A\&A}
\def\aapr{A\&A~Rev.}
\def\aaps{A\&AS}
\def\mnras{MNRAS}
\def\nat{Nature}
\def\pasp{PASP}
\def\aplett{Astrophys.~Lett.}

\bibliographystyle{mn2e}
\bibliography{references}

\appendix

\section{Models of the galaxy luminosity function at redshift 2.58}
\label{sec:LF}
In general, the observational luminosity functions are well modeled by the
\cite{S76} law, $\Phi(L) = dN/dL = \Phi^{\star} L^{\alpha} 
\exp(-L)$, where $L$ is expressed in units of a characteristic luminosity
$L^{\star}$. \cite{sha01} find, at $z\sim3$, $\Phi^{\star} = 
(6.2\pm2.7)\cdot10^{-4}$ Mpc$^{-3}$, $\alpha = -1.85\pm0.15$ and $M_V^{\star} 
= -22.98\pm0.25$. \cite{m07} find, at $z\sim3$, $\Phi^{\star} =
(16.3\pm4.6)\cdot10^{-4}$ Mpc$^{-3}$, $\alpha = -1.12\pm0.24$ and $M_V^{\star} 
= -22.77\pm0.22$. Finally, Marchesini et al. find, at $z\sim2.25$, 
$\Phi^{\star} = (11.6\pm2.9)\cdot10^{-4}$ Mpc$^{-3}$, $\alpha = -1.02\pm0.21$ 
and $M_R^{\star} = -22.67\pm0.21$. The LF of Shapley et al. is based mainly
on LBGs (Lyman Break Galaxies), and is characterized by a steep faint end
slope \citep[e.g.,][]{SLF08}. On the other hand, the LFs of
Marchesini et al. are based on LBG samples, supplemented by DRG (distant red
galaxies) samples. The faint end slopes of these LFs are considerably less
steep than for the Shapley et al. The main reason for choosing these LFs is
a desire to span a wide range of LFs in the analysis.

\begin{figure}
\includegraphics[width=0.48\textwidth]{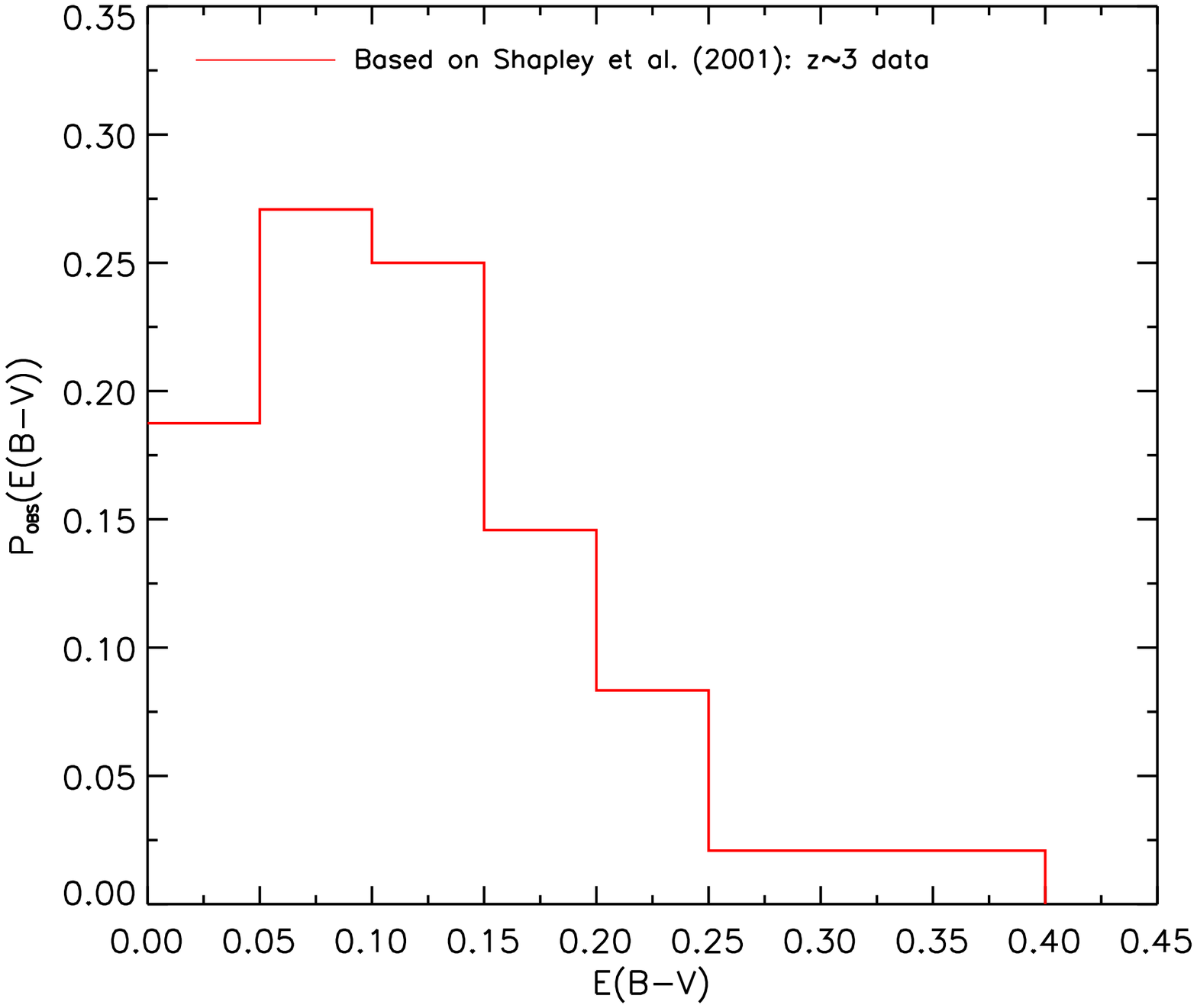}
\caption{Observationally determined E(B-V) probability distribution, based on 
a subset of the Shapley et al. data. Note that, for illustrative purposes 
only, the probability distribution has not been appropriately normalized.
\label{fig:pebv}}
\end{figure}
The above LFs are obviously {\it observational}, and it is well known that
dust extinction plays a significant role in reducing the amount of light
emitted by young galaxies, especially in the UV, e.g., \cite{RS04}, 
\cite{D.07}. Since the extinction correction to the (UV based) co-moving star 
formation density is about a factor of 5.5 \citep[e.g.,][]{SLF08}, it is 
clear that even for (rest-frame) optical bands, like V and R, the correction
is considerable. To estimate the extinction correction, we use the data of
Shapley et al., who list values of E(B-V) for 72 galaxies. They find no 
correlation between {\it observed} $M_V$ and E(B-V). In Fig.~\ref{fig:pebv}
we show the
probability distribution P$_{obs}$(E(B-V)), based on the Shapley et al. data.
Shapley et al. derive the E(B-V)s using population synthesis modeling. However,
a number of their (bright) galaxies are assigned very young ages, typically
a few tens of Myr. Such object are not seen in the simulations, so to be
conservative we base the P$_{obs}$(E(B-V)) model on the 48 (out of 72) S01 
galaxies of ages larger than 200 Myr --- the simulated galaxies typically have 
mean ages of 0.5-1 Gyr.  

\begin{figure}
\includegraphics[width=0.48\textwidth]{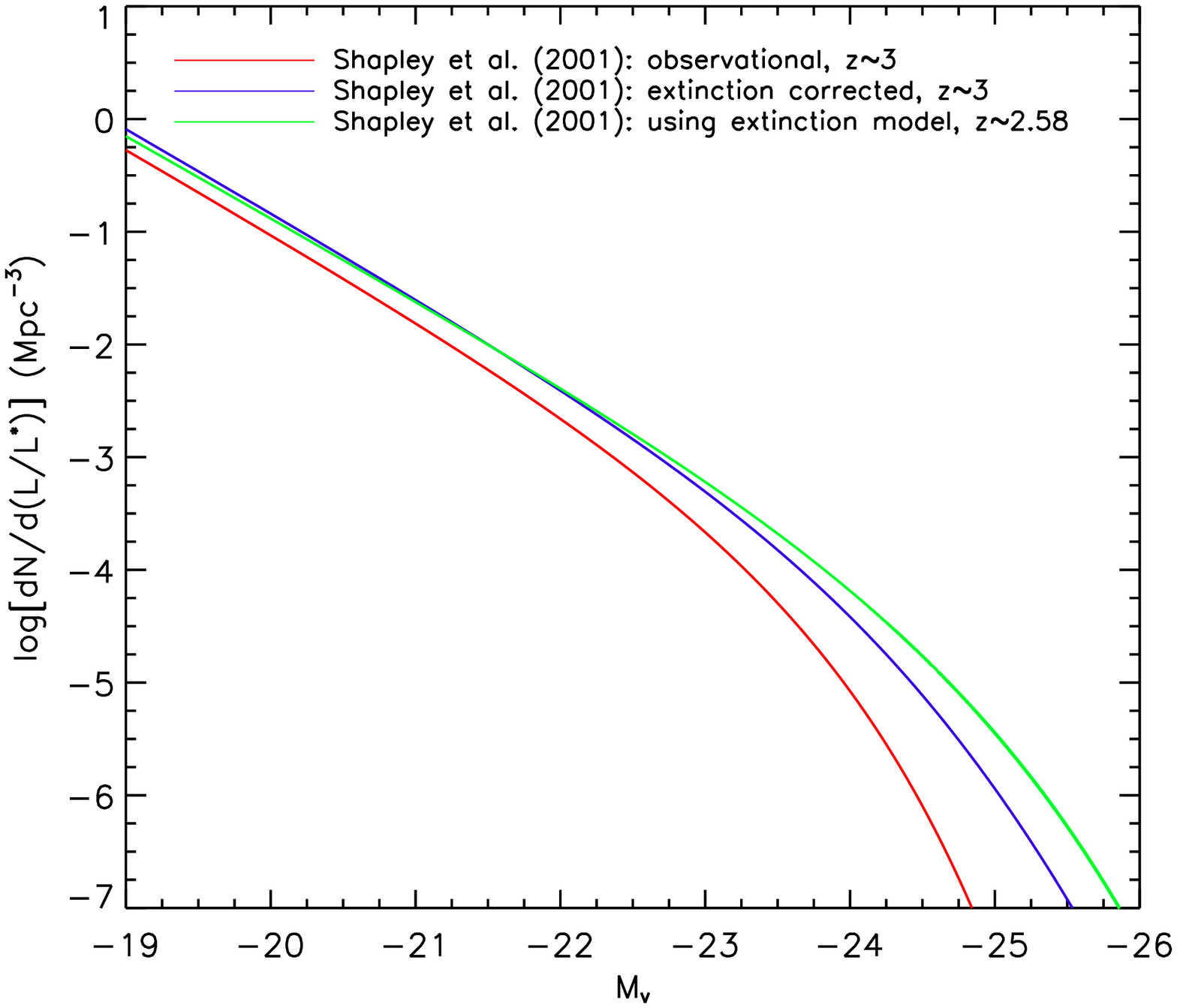}
\caption{The observational $z\sim3$ LF of Shapley et~al.~(2001) is shown by
the red curve,
together with the extinction corrected $z\sim3$ LF, obtained as described
in the text (blue curve), and with an extinction correction appropriate at 
$z$=2.58, as also described in the text (green curve).
\label{fig:LFcorr}}
\end{figure}

\begin{figure}
\includegraphics[width=0.48\textwidth]{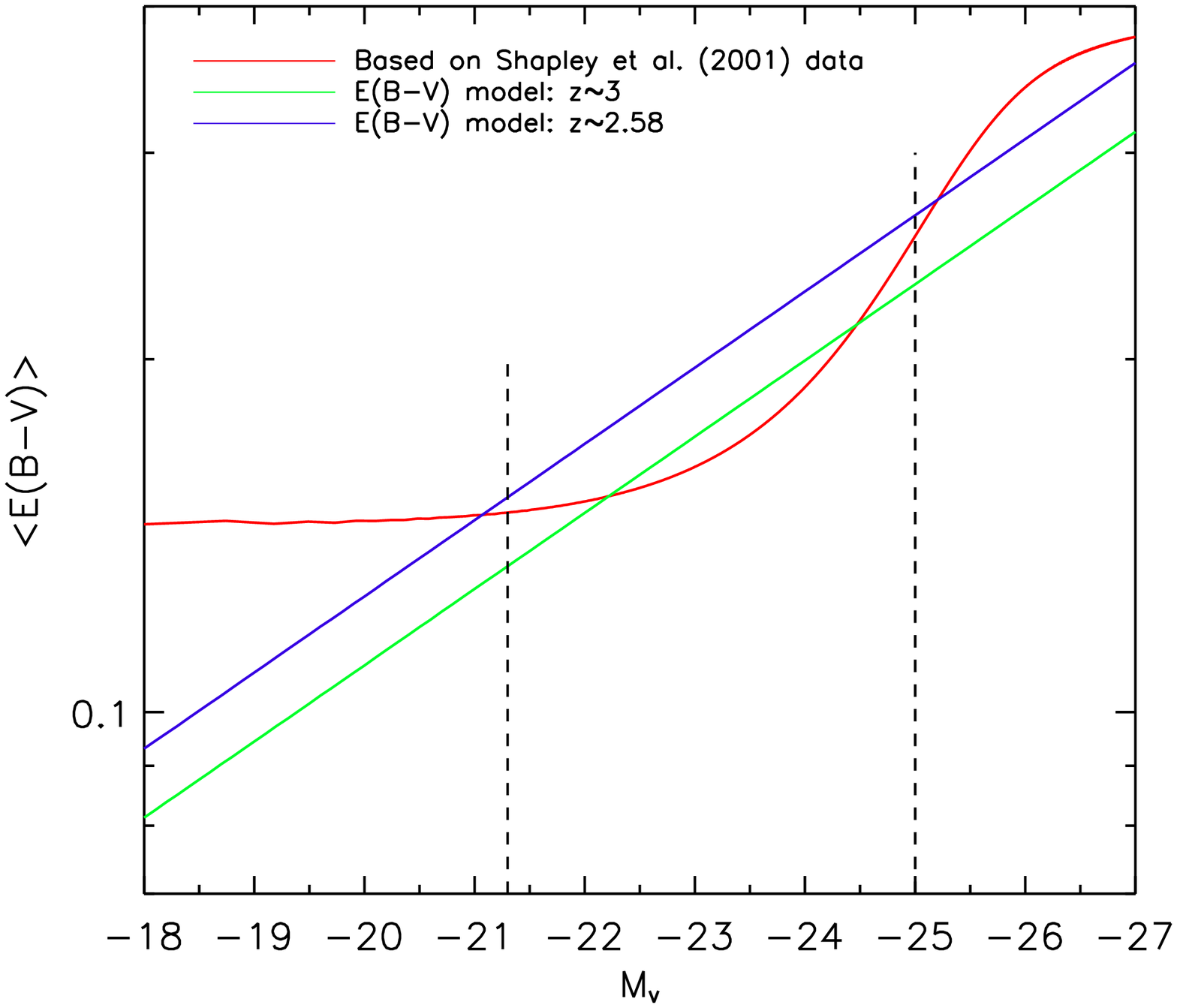}
\caption{Average E(B-V) as a function of extinction corrected $M_V$.
\label{fig:ebvmv}}
\end{figure}

In the following, the S01 observed LF is folded with the observed E(B-V)
probability distribution, under the simplifying assumption that the
observed $M_V$ and observed E(B-V) are uncorrelated, as indicated by the S01 
data. The relation between observed and de-reddened absolute V-band magnitude
is given by    
\begin{equation}\label{eq:M_V}
M_V = M_{V,obs} - A_V = M_{V,obs} - 4.05 E(B-V) ~~, 
\end{equation}
adopting a Calzetti dust law \citep{C.00}. From the definition of
magnitude, this implies that the relation between the corresponding 
luminosities is given by
\begin{equation}\label{eq:L_V}
L_V = L_{V,obs} 10^{4.05~E(B-V)/2.5} = L_{V,obs} 10^{1.62 \epsilon} ~~, 
\end{equation}
where, for brevity, E(B-V) is denoted by $\epsilon$. Alternatively, the
$\epsilon$ that connects observed luminosity $L_{obs}$ with intrinsic 
luminosity $L$ (with $L \ge L_{obs}$), must be given by
\begin{equation}\label{eq:eps}
\epsilon = \log(\frac{L_V}{L_{V,obs}})/1.62 ~~. 
\end{equation}
For a given $L_{V,obs}$, the number density of galaxies with observed 
luminosities
in the small interval [$L_{V,obs}$,$L_{V,obs}+dL_{V,obs}$] is given by
$dN = \frac{dN}{dL_{V,obs}} dL_{V,obs}$. The fraction of these that have
intrinsic luminosities in the interval [$L_V$,$L_V+dL_V$] is given by
\begin{multline}\label{eq:df}
df = P_{obs}(\epsilon)~d\epsilon = P_{obs}(\epsilon(\frac{L_V}{L_{V,obs}}))
\frac{d\epsilon}{d L_V} d L_V = \\
P_{obs}(\epsilon(\frac{L_V}{L_{V,obs}})) \frac{d L_V}{1.62~\ln(10)~L_V} ~~,
\end{multline}
where eq.~(\ref{eq:eps}) has been used. Hence, the number density of galaxies 
with observed luminosities in the small interval of size $dL_{V,obs}$ and 
intrinsic luminosities in the small interval of size $dL_V$, can be expressed
as
\begin{equation}\label{eq:d2n}
d^2 N = \frac{dN}{dL_{V,obs}} \cdot
\frac{P_{obs}(\epsilon(\frac{L_V}{L_{V,obs}}))}{1.62~\ln(10)~L_V}~dL_{V,obs}~
dL_V ~~.
\end{equation}
The extinction corrected LF is the obtained by integration over $L_{V,obs}$:
\begin{equation}\label{eq:int}
\frac{dN_{corr}}{dL_V}(L_V) = \int_0^{L_V}\frac{dN}{dL_{V,obs}} \cdot
\frac{P_{obs}(\epsilon(\frac{L_V}{L_{V,obs}}))}{1.62~\ln(10)~L_V}~dL_{V,obs} 
~~.
\end{equation}
We now change variable in the integral in eq.~(\ref{eq:int}). It follows from
eq.~(\ref{eq:L_V}) that $L_{V,obs} = L_V \cdot 10^{-1.62 \epsilon}$, and
hence that $dL_{V,obs}/d\epsilon = -1.62~\ln(10)~L_V~10^{-1.62 \epsilon}$.
Eq.~(\ref{eq:int}) can then be rewritten as
\begin{multline}\label{eq:int1}
\frac{dN_{corr}}{dL_V}(L_V) = \\
\int_{\epsilon_{max}}^0\frac{dN}{dL_{V,obs}}
(L_{V,obs}) \cdot 
\frac{dL_{V,obs}}{d\epsilon} \cdot
\frac{P_{obs}(\epsilon)}{1.62~\ln(10)~L_V}~d\epsilon \\
= \int_0^{\epsilon_{max}}\frac{dN}{dL_{V,obs}}(10^{-1.62 \epsilon}L_V)
P(\epsilon)~10^{-1.62 \epsilon}~d\epsilon ~~,
\end{multline}
where $\epsilon_{max}$ is the maximum observed value of E(B-V) - based on the
data of S01, we set $\epsilon_{max}$=0.4 (see Fig.~\ref{fig:pebv}).
Fig.~\ref{fig:LFcorr} shows the result of applying the above procedure
to the observational S01 $z\sim3$ LF, together with the original S01 LF itself.
If the observational LF is described by a power-law (i.e., scale-free), then,
if the above approach is used, the extinction corrected LF will be a power-law
with the same power index, just shifted towards larger $L_V$. Due to the
``knee'' of the luminosity function (caused by the exponential cut-off), 
however, the shapes of the observational and corrected LFs will be different,
as can be seen from the Figure. In the same way, one can show that if 
the observational LF is described by a power-law, then the average E(B-V) of
galaxies will be independent of uncorrected luminosity. This seems at odds
with the fact that galaxy metallicity is correlated with luminosity, and
it is typically assumed that the dust content in galaxies is correlated with
the metallicity \citep[e.g.,][]{L.09}. Due to
the ``knee'' of the observational LF, however, extinction corrected luminosity 
is correlated with E(B-V), as one would expect --- see Fig.~\ref{fig:ebvmv}, 
and also S01. As can also be seen from the figure, $<$E(B-V)$>$ approaches
a constant value for luminosities fainter than $M_V \ga$ -22 to -21. This
seems not physically reasonable (see above), and is a consequence of the
assumption that E(B-V) is uncorrelated with observed luminosity (based on
the S01 data in the limited range $M_{V,obs} \sim$-21 - -24), and the LF
approximating a power-law at lower luminosities, cf. the discussion above. 

In order to obtain a physically more reasonable extinction model, we consider
the following model, where, for simplicity, E(B-V) is modeled as a 
{\it function}
of $M_V$ with the properties that a) E(B-V) decreases monotonically with
decreasing intrinsic luminosity, and b) provides a good match to the S01
data over the limited $M_V$ range available: 
\begin{equation}\label{eq:ebvmodel}
\log(\epsilon) = -0.065(M_V+26)-0.57 ~~, ~~~~z \sim 3 ~~, 
\end{equation}
where E(B-V), as before, is denoted $\epsilon$. As can be seen from 
Fig.~\ref{fig:ebvmv}, the model provides a good match to the S01 data, in
the limited range of $M_V$ available.

The S01 LF is built on observations of $z \sim 3$ galaxies, so the above
E(B-V) model obviously pertains to such a redshift. The DLA under discussion in
this work is located at $z$=2.58, and the amount of heavy elements in galaxies
will have increased during this $\sim$400 Myr period. So when using the S01 
observational LF model also as an observational LF at $z$=2.58, it is likely
that a somewhat larger dust extinction correction is required to obtain
the extinction corrected LF. To quantify this, we determined the metallicity
of all simulated galaxies in this project, at both $z$=3 and $z$=2.58. We
find an average increase in abundance at a given $M_V$ of 0.06 dex or
equivalently 15\%. It reasonable to assume that the average dust opacity
has increased by a similar amount \citep[e.g.,][]{L.09}, 
and, adopting a simple ``screen'' model, so has the average E(B-V).
The final E(B-V) model adopted therefore reads
\begin{equation}\label{eq:ebvmodel2.58}
\log(\epsilon) = -0.065(M_V+26)-0.51 ~~, ~~z \sim 2.58 ~~. 
\end{equation}
This model is also shown in Fig.~\ref{fig:ebvmv}. In this paper, the
extinction corrected, $z$=2.58 LFs are determined on the basis of this 
model --- see also below. It must hold that
\begin{equation}\label{eq:lfmodel}
\frac{dN_{corr}}{dL_V}(L_V) = \frac{dN}{dL_{V,obs}}(L_{V,obs}(L_V)) \cdot
\frac{dL_{V,obs}}{dL_V} ~~.
\end{equation}
It follows from eq.~(\ref{eq:L_V}) that
\begin{equation}\label{eq:L_Vobs}
\log(L_{V,obs}) = \log(L_V) - 1.62 \epsilon ~~. 
\end{equation}
Inserting the extinction model yields
\begin{equation}\label{eq:L_Vobs1}
\log(L_{V,obs}) = \log(L_V) - 1.62 \cdot 10^{-0.065(M_V+26)-0.51} ~~. 
\end{equation}
Now, since in this work luminosities are in expressed in units of 
$L^{\star}$, it follows from S01, that $M_V = -22.98-2.5 \log(L_V)$~.
Inserting this in eq.~(\ref{eq:L_Vobs1}) yields
\begin{equation}\label{eq:L_Vobs2}
\log(L_{V,obs}) = \log(L_V) - 0.319 L_V^{0.163} ~~, 
\end{equation}
which can be rewritten as
\begin{equation}\label{eq:L_Vobs3}
L_{V,obs} = L_V \cdot 0.48^{L_V^{0.163}} ~~. 
\end{equation}
Taking the derivative of this equation results in
\begin{equation}\label{eq:L_Vobs4}
\frac{dL_{V,obs}}{dL_V} = 0.48^{L_V^{0.163}}(1.0 - 0.119~L_V^{0.163}) ~~. 
\end{equation}
Using eq.~(\ref{eq:lfmodel}), the final 
extinction corrected LF model can be expressed as
\begin{multline}\label{eq:lfmodel1}
\frac{dN_{corr,1}}{dL_V}(L_V) = \\
\Phi^{\star} (0.48^{L_V^{0.163}}L_V)^{\alpha} 
\exp(-0.48^{L_V^{0.163}}L_V) (1.0 - 0.119~L_V^{0.163}) ~~,
\end{multline}
where $\Phi^{\star} = 6.2\cdot10^{-4}$ Mpc$^{-3}$, and $\alpha=-1.85$ This LF 
model is also shown in Fig.~\ref{fig:LFcorr}. As can be seen from the Figure, 
this LF is similar to the $z\sim3$ one obtained by extinction correcting
using the probabilistic approach 
described above (eq.~(\ref{eq:int1})), though at $M_V \la -23$ the $z=2.58$ LF 
model lies slightly above --- this is due to the 15\% larger dust opacity,
as well as other differences between the two types of extinction corrections.
For the purposes of this paper, the $z$=2.58 extinction corrections based on
the probabilistic approach and on the extinction model (\ref{eq:ebvmodel2.58})
lead to almost identical results --- in the this paper, model 
(\ref{eq:ebvmodel2.58}) has been used exclusively.

\begin{figure}
\includegraphics[width=0.48\textwidth]{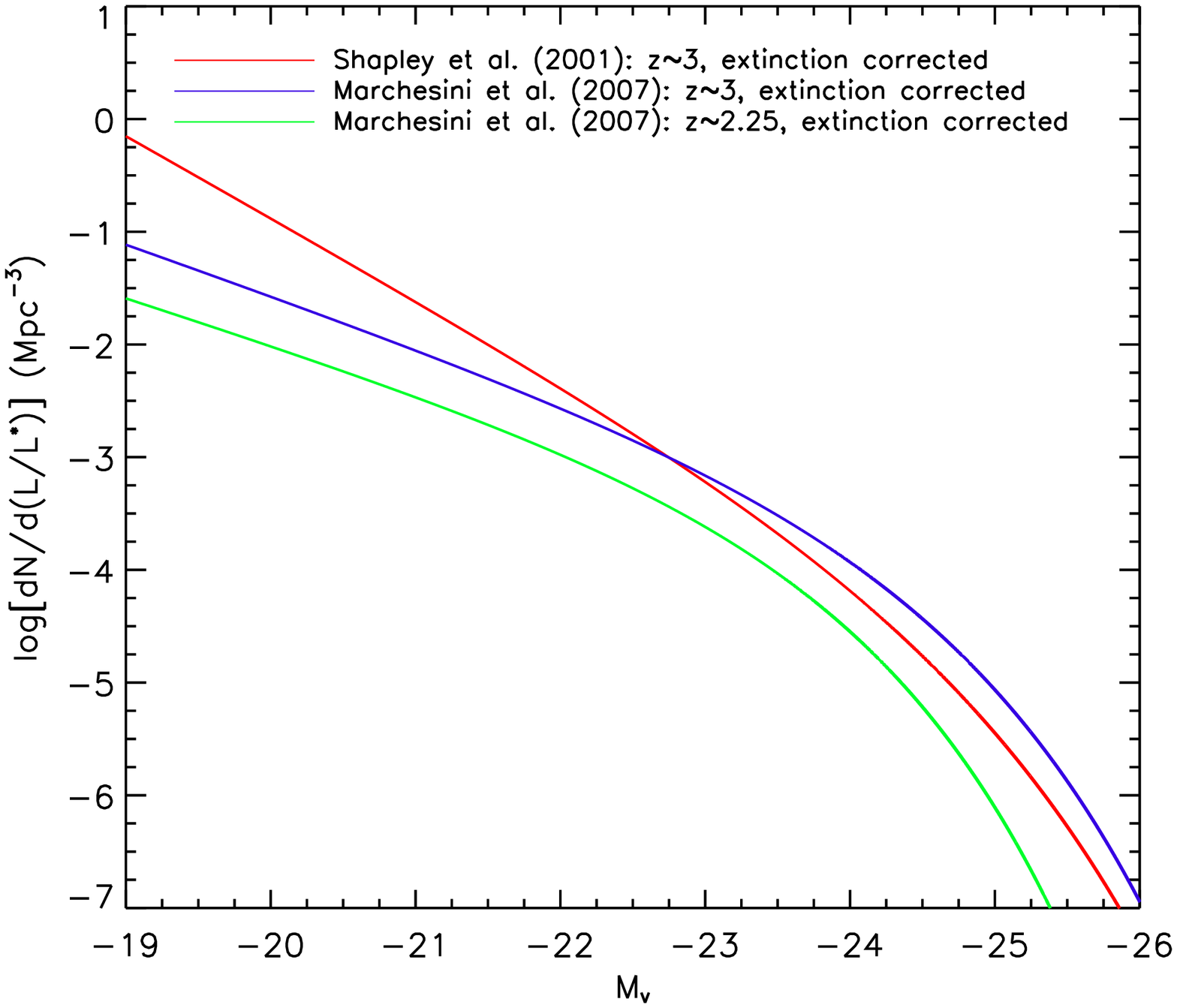}
\caption{The three extinction corrected LFs, corrected using the $z=2.58$
extinction model (eq.~(\ref{eq:ebvmodel2.58})).
\label{fig:LFfinal}}
\end{figure}

The M07$z$$\sim$3 LF is also based on rest-frame V-band observations, and 
the extinction correction using the $z=2.58$ extinction model 
(eq.~(\ref{eq:ebvmodel2.58})) follows the same procedure as outlined above.
The resulting extinction corrected LF is given by
\begin{multline}\label{eq:lfmodel2}
\frac{dN_{corr,2}}{dL_V}(L_V) = \\
\Phi^{\star} (0.491^{L_V^{0.163}}L_V)^{\alpha} 
\exp(-0.491^{L_V^{0.163}}L_V) (1.0 - 0.116~L_V^{0.163}) ~~,
\end{multline}
where $\Phi^{\star} = 16.3\cdot10^{-4}$ Mpc$^{-3}$, and $\alpha=-1.12$ This LF 
model is also shown in Fig.~\ref{fig:LFfinal}.
The M07$z$$\sim$2.25 LF is based on rest-frame R-band observations, and it is
slightly more elaborate to perform the extinction correction using the 
$z=2.58$ extinction model (eq.~(\ref{eq:ebvmodel2.58})), which is based
on the V-band absolute magnitude. We proceed as follows: For the simulated
galaxies at $z=2.58$, we find an average V-R of about 0.25 mag. For the 
purposes of the present work, we simply set $M_V = M_R + 0.25$. Substituting
this into eq.~(\ref{eq:ebvmodel2.58}) yields
\begin{equation}\label{eq:ebvmodel2.58R}
\log(\epsilon) = -0.065(M_R+26.25)-0.51 ~~, ~~z \sim 2.58 ~~,
\end{equation}
so E(B-V) is now expressed in terms of $M_R$. For Calzetti dust \citep{C.00}
\begin{equation}\label{eq:M_R}
M_R = M_{R,obs} - A_R = M_{R,obs} - 3.33 E(B-V) ~~.
\end{equation}
Hence, by the definition of absolute magnitude it follows that
\begin{equation}\label{eq:L_Robs}
\log(L_{R,obs}) = \log(L_R) - 1.33 \epsilon ~~. 
\end{equation}
Inserting eq.~(\ref{eq:ebvmodel2.58R}) in this yields
\begin{equation}\label{eq:L_Robs1}
\log(L_{R,obs}) = \log(L_R) - 1.33 \cdot 10^{-0.065(M_R+26.25)-0.51}  ~~. 
\end{equation}
Now, as luminosities are in expressed in units of $L^{\star}$, it follows 
that $M_R = -22.67-2.5 \log(L_R)$ at $z\sim2.25$ \citep{m07}. 
Inserting this yields
\begin{equation}\label{eq:L_Robs2}
\log(L_{R,obs}) = \log(L_R) - 0.24 L_R^{0.163} ~~, 
\end{equation}
which can be rewritten as
\begin{equation}\label{eq:L_Robs3}
L_{R,obs} = L_R \cdot 0.575^{L_R^{0.163}} ~~. 
\end{equation}
Taking the derivative of this equation results in
\begin{equation}\label{eq:L_Robs4}
\frac{dL_{R,obs}}{dL_R} = 0.575^{L_R^{0.163}}(1.0 - 0.09~L_R^{0.163}) ~~. 
\end{equation}
Finally, using the R-band equivalent of eq.~(\ref{eq:lfmodel}), the final 
extinction corrected LF model can be expressed as
\begin{multline}\label{eq:lfmodel3}
\frac{dN_{corr,3}}{dL_R}(L_R) = \\
\Phi^{\star} (0.575^{L_R^{0.163}}L_R)^{\alpha} 
\exp(-0.575^{L_R^{0.163}}L_R) (1.0 - 0.09~L_R^{0.163}) ~~,
\end{multline}
where $\Phi^{\star} = 11.6\cdot10^{-4}$ Mpc$^{-3}$, and $\alpha=-1.02$. 
This LF model is also shown in Fig.~\ref{fig:LFfinal} --- for comparison
with the two other LF models, it is shown as a function of $M_V$, using
$M_V = M_R +0.25$ (see above).

\end{document}